\documentclass[10pt]{bmc_article}
\usepackage{amsmath,amsfonts}
\usepackage{amssymb,amsbsy}
\usepackage{graphics,epsfig}

\newcommand\ignore[1]{}
\newcommand{\argmax}{\operatornamewithlimits{arg\ max}}

\newcommand{\hspp}{\hspace{0.05in} }
\newcommand{\hsppp}{\hspace{0.02in} }

\newcommand{\indic}{\mbox{$1\!\!1$}}

\usepackage{cite} 
\usepackage{url}  
\usepackage{ifthen}  
\usepackage{multicol}   
\usepackage[utf8]{inputenc} 
\newboolean{publ}

\newenvironment{bmcformat}{\baselineskip20pt\sloppy\setboolean{publ}{false}}{\baselineskip20pt\sloppy}

\begin{document}
\begin{bmcformat}


\title{Modeling Temporal Activity Patterns in Dynamic Social Networks}


\author{Vasanthan Raghavan\correspondingauthor$^1$,%
         \email{Vasanthan Raghavan\correspondingauthor - vasanthan\_raghavan@ieee.org}
         Greg Ver Steeg$^2$,
         \email{Greg Ver Steeg - gregv@isi.edu}
         Aram Galstyan$^2$, and
         \email{Aram Galstyan - galstyan@isi.edu}Alexander G. Tartakovsky$^1$
         \email{Alexander G. Tartakovsky - tartakov@usc.edu}
               }


\address{%
    \iid(1)Department of Mathematics, University of Southern California, Los
    Angeles, 90089, CA, USA \\
    \iid(2)Information Sciences Institute, University of Southern California,
    Marina del Rey, 90292, CA, USA }%

\maketitle

\begin{abstract}
\noindent
The focus of this work is on developing probabilistic models for user activity
in social networks by incorporating the social network {\em influence} as perceived
by the user. For this, we propose a {\em coupled} Hidden Markov Model, where each
user's activity evolves according to a Markov chain with a hidden state that is
influenced by the collective activity of the friends of the user. We develop
generalized Baum-Welch and Viterbi algorithms for model parameter learning and
state estimation for the proposed framework. We then validate the proposed
model using a significant corpus of user activity on {\tt Twitter.} Our numerical
studies show that with sufficient observations to ensure accurate model learning,
the proposed framework explains the observed data better than either a renewal
process-based model or a conventional uncoupled Hidden Markov Model. We also
demonstrate the utility of the proposed approach in predicting the time to the
next tweet. Finally, clustering in the model parameter space is shown to result
in distinct natural clusters of users characterized by the interaction dynamic
between a user and his network.
\end{abstract}

\ifthenelse{\boolean{publ}}{\begin{multicols}{2}}{}

\section*{Keywords}
Activity Profile Modeling, {\tt Twitter}, Data-Fitting, Explanation, Prediction,
Hidden Markov Model, Coupled Hidden Markov Model, Social Network
Influence, User Clustering



\section*{Introduction}
\label{sec1}
Social networking websites such as ${\tt Facebook}$, ${\tt Twitter}$, etc.\
have become immensely popular with hundreds of millions of users that engage
in various forms of activity on these websites. These social networks provide an
unparalleled opportunity to study individual and collective behavior at a
very large scale. Such studies have profound implications on wide-ranging
applications such as efficient resource allocation, user-specific information
dissemination, user classification, and rapid detection of anomalous behavior
such as bot or compromised accounts, etc.

\ignore{
Another important problem that has attracted significant interest
is characterizing individual and collective activity patterns in such settings.
Understanding temporal patterns of user activity can be leveraged for a number
of important applications, such as
}

The simplest model for user activity is a Poisson process, where each activity
event (e.g., posting, tweeting, etc.) occurs independent of the past history at
a time-independent rate. However, recent empirical evidence from multiple
sources (e-mail logs, web surfing, letter correspondence, research output, etc.)
suggest that human activity has distinctly non-Poissonian characteristics. In
particular, the inter-event duration distribution (which is exponential for the
Poisson process) has been shown to be heavy-tailed and bursty for a number of
different activity types~\cite{barabasi05,goh_barabasi}. Different approaches
have been put forward to explain the non-Poisson nature of the activity patterns~\cite{vazquez06,vazquez07,malmgren_pnas,malmgren1,malmgren,Bianconi,Rybski,jo_2012}.
Nevertheless, despite significant recent progress, open questions remain. Most
remarkably, on an individual scale, existing studies so far have mostly discarded
the role and impact of the social network where the user activity takes place,
instead describing each user via an {\em independent} stochastic process. On the
other hand, it is clear that social interactions on networks affect user activity,
and discarding these interactions should generally lead to sub-optimal models.

The main contribution of this paper is to develop a probabilistic model of
user activity that explicitly takes into account the interaction between users
by introducing a coupling between two stochastic processes. Specifically, we
propose a coupled Hidden Markov Model (coupled HMM) to describe inter-connected
dynamics of user activity. In our model, the individual dynamics of each user
is coupled to the aggregated activity profile of his neighbors (friends or
followers) in the network. While a user's activity may be preferentially affected
by specific neighbors, the predictive power of the model can be substantially
improved using the aggregated activity of all the neighbors. The hidden states
in our model correspond to different patterns in user activity, similar to the
approach suggested in~\cite{malmgren}. However, here the state transitions are
influenced by the activity of the neighbors, and in turn, the activity of the
aggregated set of neighbors is influenced by the state of the given user. While
many variants of the classical HMM approach exist in the literature, the key
distinguishing feature of this work is the {\em bi-directional} influence between
the activity of a user and his network, specifically tailored for social network
applications. Nevertheless, being a variant of the conventional (uncoupled) HMM,
the proposed model enjoys the same computational advantages in terms of parameter
learning as~\cite{malmgren} and other HMM variants. 

We perform a number of experiments with data describing user activity traces
on ${\tt Twitter}$, and demonstrate that the proposed approach has a better
performance both in terms of explaining observed data (model fitting) and
predicting future activity (generalization). In particular, we report
statistically significant improvement over two baseline approaches, a renewal
process-based model and a conventional HMM. Furthermore, we use the
learned models to cluster users, and find that the resulting cluster structure
allows intuitive characterization of the users in terms of the interaction
dynamics between a user and his social network.

The rest of the paper is organized as follows. Section~2 reviews related
work in activity profile modeling, especially as applicable in social network
settings. Section~3 proposes a coupled HMM framework for the activity profile
of users and distinguishes the proposed model from prior work. Section~4
develops statistical methodologies to learn model parameters and how to use
the proposed framework for data-fitting and forecasting. Section~5 validates
the modeling assumptions and illustrates the utility and efficacy of the
proposed approach with data from~\cite{sofus}. Concluding remarks are provided
in Section~6. Without any prejudice, we use the male gender-specific connotations
for all the users and their attributes in this work.

\section*{Related Work}
\label{sec2}
Recent research on social networks has focused on understanding the properties
of networks induced by social interactions, modeling information diffusion on
such networks, characterizing their evolution in time, etc. In the direction of
modeling the temporal activity of users' communication in such networks, several
models have been proposed in the literature. Approaches based on simple point
process models have been proposed for user requests in a peer-to-peer
setting~\cite{guo_chen_sigcomm2005} and social network
evolution~\cite{leskovec_sigkdd2008}. In addition to one-parameter exponential
observation density for user activity utilized in~\cite{malmgren1}, more general
two-parameter models such as the Weibull (or stretched exponential) have been
proposed for modeling inter-post duration in the context of instant-messaging networks~\cite{xiao_guo_icdcs2007}, accessing patterns in
Internet-media~\cite{guo_tan_podc2008}, understanding inter-post dynamics of
original content in general online social networks~\cite{guo_tan_kdd2009}, and
inter-call duration in cell-phone networks~\cite{jiang}.

To explain the bursty features of human dynamics,~\cite{barabasi05,goh_barabasi,vazquez06,vazquez07} suggested the
priority selection/queue mechanism. An alternative mechanism motivated by
circadian and weekly cycles of human activity and captured by cascading
non-homogenous Poisson processes was suggested in~\cite{malmgren_pnas,malmgren1}
for the heavy-tails in inter-event durations. Although this model has been
shown to be consistent with empirical observations, it is computationally
intensive in terms of parameter estimation. To overcome this issue, Malmgren
{\em et al.}~\cite{malmgren} suggested a simpler two-state HMM for the activity
of users in an email/communication network where the states reflect a measure
of the user's activity. Similar models, suggestive of a few states determining
activity patterns, have been considered for short message correspondence
in~\cite{wu_zhou} and Digg activity in~\cite{hogg_lerman}. Other work has also
emphasized the importance of distinguishing active versus inactive users in activity
and influence modeling tasks~\cite{Romero2010}.

An important feature that characterizes the priority selection mechanism is
that human behavioral patterns are driven by responses to/from others. On the
other hand, the line of work motivated by the cascading non-homogenous
Poisson processes-based models explains activity patterns due to other
mechanisms such as circadian and weekly cycles, task repetition, and
changing communication needs. Models that traverse between these two extremes
and incorporate the influence of a user's social network on activity have
not been studied in extensive detail in the literature. Some of the examples
of such a bridging effort include user-generated activity traces used for
inferring underlying social relationships~\cite{GomezRodriguez2010,zhao,VerSteeg2012WWW},
incorporating similarities in the behavior of users in a specific user's
social network to model his actions~\cite{jie_tang2010,jie_tang2011},
a Poisson regression model to determine the users that most influence a
given user~\cite{trusov}, and a user activity-driven (rather than
connectivity-driven) model for social network evolution~\cite{perra}.
Notwithstanding the fact that similar ideas have been sporadically pursued in
other settings, our work provides the missing link for a network-driven
approach to capturing individual behavioral patterns.

While the theory of classical HMMs is well-developed~\cite{rabiner,bilmes},
HMMs are ill-suited in settings where multiple processes interact with each
other and/or information about the history of the process needed for future
inferencing is not reflected in the current state. To incorporate complicated
dependencies between several interacting variables, many variants of the
classical HMM set-up that are special cases of the more general theory of
{\em dynamic Bayesian networks}~\cite{pearl,murphy_diss} have been introduced
in the literature. Some of these extensions include the class of
{\em autoregressive} HMMs~\cite{poritz,juang_rabiner1,kenny_lennig},
{\em input-output} HMMs~\cite{Cacciatore94mixturesof,bengio},
{\em factorial} HMMs~\cite{ghahramani}, and {\em coupled} HMMs. Different
variants of coupled HMMs have been used in diverse settings including models
for complex human actions and behaviors~\cite{brand_oliver_pentland,pavlovic},
freeway traffic~\cite{kwon_murphy}, audio-visual speech~\cite{chu_huang1,chu_huang2},
EEG classification~\cite{zhong_ghosh}, spread of infection in social
networks~\cite{dong_pentland}, etc.

\section*{Modeling Activity Profile of Twitter Users}
\label{sec3}
Let $T_{i}, \hsppp i = 0, 1, \cdots, N$ denote the time-stamps of a
specific user's tweets over a given period-of-interest. We can equivalently
define the inter-tweet duration $\Delta_i$ as
\begin{eqnarray}
\Delta_{i} \triangleq T_{i} - T_{i-1}, \hspp i = 1, 2, \cdots , N.
\nonumber
\end{eqnarray}
One of the main goals of this work is to develop a mathematical model
for $\{ \Delta_i \} \triangleq \Delta_1^N = \left[ \Delta_1, \cdots, \Delta_N
\right]$. Along the lines of~\cite{malmgren}, we start by developing a
simplistic $k = 2$-state HMM for $\{ \Delta_i \}$.

\subsection*{Influence-Free Hidden Markov Modeling}
\label{sec3a}

\noindent
{\bf \em Assumption 1 -- Underlying States:} We assume that a variable
$Q_i$, taking one of two possible values $\{ 0, 1\}$, reflects the state of
the user-of-interest. Specifically, $Q_i = 0$ denotes that the user is in an
{\em Inactive} state between $T_{i-1}$ and $T_i$, whereas $Q_i = 1$ denotes
that the user is in an {\em Active} state. We also assume that
$Q_i \hspp (i \geq 1)$ evolves in a time-homogenous Markovian manner and
is dependent only on $Q_{i-1}$ and is conditionally independent of
$Q_0^{i-2} = \left[ Q_0, \cdots, Q_{i-2} \right]$ given $Q_{i-1}$. A
first-order Markovian model is a reasonable first approximation capturing
short-term memory in human behavioral dynamics~\cite{goh_barabasi,karsai}.
The state transition probability matrix ${\sf P} = \{ {\sf P}[m,n] \}$ is
given as
\begin{eqnarray}
{\sf P} = \left[
\begin{array}{cc}
1 - \beta_{0,1} & \beta_{0,1} \\
\beta_{1,0} & 1 - \beta_{1,0} \end{array}
\right],
\nonumber
\end{eqnarray}
with ${\sf P}[m,n] = {\sf P}(Q_i = n | Q_{i-1} = m), \hspp m, n
\in \{ 0,1 \}$. The density of the initial state $Q_0$ is denoted as
${\sf P}( Q_0 = j) = \pi_j, \hspp j = 0,1$. Note that the switching
from the {\em Inactive} state to the {\em Active} state in the HMM
paradigm can capture the nocturnal/work-home patterns of individual
users without any further explicit modeling~\cite{malmgren}. Further,
explicit modeling of circadian and weekly cycles in social network
settings is more difficult than for email communications due to the
``often on'' more random nature of social network interactions.

\noindent {\bf \em Assumption 2 -- Observation Density:} In general,
$Q_i$ is hidden (unobservable) and we can only observe $\{ \Delta_i \}$
(or equivalently, $\{ T_i \}$). In the {\em Inactive} state,
$\{ \Delta_i \}$ form samples from a ``low''-rate point process,
whereas in the {\em Active} state, $\{ \Delta_i \}$ form samples
from a ``high''-rate point process. Specifically, let the probability
density function of $\Delta_i$ be given as
\begin{eqnarray}
\Delta_i \sim \left\{
\begin{array}{cc}
f_1( \cdot ) & {\rm if} \hspp \hspp Q_i = 1 \\
f_0( \cdot ) & {\rm if} \hspp \hspp Q_i = 0,
\end{array}
\right.
\nonumber
\end{eqnarray}
for an appropriate choice of $f_0(\cdot)$ and $f_1(\cdot)$.

As mentioned earlier, an exponential model for $f_{\cdot}(\cdot)$
corresponds to a Poisson process assumption under either state. While
the exponential model is captured by a single parameter (see
Table~1 
for details), this simplicity often constrains the
model fit either in the small inter-tweet (bursty) regime or large
inter-tweet regime (tails). Two-parameter extensions of the exponential
such as the gamma or Weibull density allow a better fit in these two
regimes. Both the gamma and the Weibull models result in similar modeling
fits. While the gamma model allows for simple parameter estimate formulas
(see Table~1), 
the Weibull model requires the solving of coupled
equations in the model parameters (often, a numerically intensive procedure).
Thus, we will restrict attention to the exponential and gamma model choices
in this work.

\subsection*{Influence-Driven Hidden Markov Modeling}
\label{sec3b}
A more sophisticated influence-driven model is developed now
by making the following additional assumptions:

\noindent
{\bf \em Assumption 3 -- Influence of Neighbors:} In addition to
$Q_{i-1}$, the evolution of $Q_i$ is also influenced by the aggregated
activity of all the users interacting with and influencing the user-of-interest
(``neighbors,'' for short). While neighbors is a broad rubric, we
restrict attention to the friends and followers in this work. For example,
a series of tweets from the neighbors can result in a reply/retweet by the
user, or a long period of non-activity from the neighbors could induce the
user to initiate a burst of activity. Let the variable $Z_i$ ($i = 1, \cdots ,
N$) capture the influence of the neighbors' tweets on the user-of-interest.
Examples of candidate influence structures (information theoretically,
$Z_i$ is viewed as a side-information metric) include:
\begin{enumerate}
\item
A binary indicator function that reflects whether there was a
mention of the user between $T_{i-1}$ and $T_i$ (or not);
\item The number of such mentions;
\item Total traffic (aggregated activity) of the friends of
the user, etc. 
\end{enumerate}

Motivated by the same short-term memory assumption as before, the
{\em coupling} between $\{Q_i\}$ and $\{ Z_i\}$ is simplified by the
Markovian condition that ${\sf P}(Q_i | Q_1^{i-1}, \hsppp Z_1^i ) =
{\sf P}(Q_i | Q_{i-1}, \hsppp Z_i)$. In general, to keep computational
requirements in inferencing low, it is helpful to assume that the evolution
of $Q_i$ is captured by a summary statistic $\phi(Z_i) \hsppp
: \hsppp Z_i \mapsto [0,1]$
such that
\begin{eqnarray}
{\sf P}(Q_i | Q_{i-1}, \hsppp Z_i) =
{\sf P}_{\sf 0}(Q_i | Q_{i-1})  \cdot \left(1 - \phi(Z_i) \right)
+ {\sf P}_{\sf 1}(Q_i | Q_{i-1})
\cdot \phi(Z_i) \nonumber
\end{eqnarray}
with ${\sf P}_{\sf k}[m,n] = {\sf P}_{\sf k}(Q_i = n | Q_{i-1} = m)$ where
\begin{eqnarray}
\begin{array}{cc}
{\sf P}_{ {\sf 0}}  =
\left[
\begin{array}{cc}
1 - p_0 & p_0 \\
q_0 & 1 -q_0
\end{array}
\right]  & {\rm and} \hspp \hspp
{\sf P}_{ {\sf 1}}  =
\left[
\begin{array}{cc}
1 - p_1 & p_1 \\
q_1 & 1 -q_1
\end{array}
\right].
\end{array}
\nonumber
\end{eqnarray}
In particular, the choice $\phi(Z_i) = \indic(Z_i > \tau)$ for a suitable
threshold $\tau$ implies that the user switches from the transition probability
matrix ${\sf P}_{\sf 0}$ to ${\sf P}_{\sf 1}$ depending on the magnitude of the
influence structure. 
To paraphrase, the user evolves according to a baseline dynamics corresponding
to ${\sf P}_0$ if his network activity is below a certain threshold and evolves
according to an elevated dynamics corresponding to ${\sf P}_1$ if his network
activity exceeds that threshold. The discussion on model learning elaborates on a
simple method to determine the appropriate choice of $\tau$. The discussion on
model validation provides some empirical justification for this assumption.

\noindent
{\bf \em Assumption 4 -- Evolution of Influence Structure:} Noting that $Z_i$ is
a function of the activity of all the neighbors (and not a specific user), we
hypothesize that $Z_i$ is dependent on $Q_{i-1}$, but only weakly dependent on
$Z_{i-1}$. Motivated by this thinking, we make the simplistic assumption that
\begin{eqnarray}
{\sf P}(Z_i | Q_{1}^{i-1}, Z_1^{i-1} ) =
{\sf P}(Z_i | Q_{i-1}). 
\label{eqn4}
\end{eqnarray}
Rephrasing,~(\ref{eqn4}) presumes that user aggregation de-correlates $Z_i$ from
its past history. While the above assumption can be justified under certain scenarios
(see the discussion under model validation), more general influence evolution models
need to be considered and the loss in explanatory/predictive power by making the
simplistic assumption in~(\ref{eqn4}) needs to be studied carefully. This is the
subject of ongoing work. Further, let the probability density function of $Z_i$ be
given as
\begin{eqnarray}
Z_i \sim \left\{ \begin{array}{cc}
g_0(\cdot) & {\rm if} \hspp \hspp Q_{i-1} = 0 \\
g_1(\cdot) & {\rm if} \hspp \hspp Q_{i-1} = 1.
\end{array}
\right.
\nonumber
\end{eqnarray}
Different candidates for $g_j(\cdot)$ are illustrated in Table~2. 

Combining the above four assumptions, the joint density of the
observations $\{ \Delta_i \}$, the influence structure $\{ Z_i \}$,
and the state $\{ Q_i \}$ can be simplified as
\begin{eqnarray}
{\sf P}  \Big( \Delta_1^N, Z_1^N, Q_0^N \Big)
& = &
{\sf P} \Big ( Q_0, \hsppp Z_1, \hsppp Q_1, \hsppp \Delta_1, \hsppp
\cdots , \hsppp Z_N, \hsppp Q_N, \hsppp \Delta_N \Big)
\\
& = & {\sf P}(Q_0 ) \prod_{i=1}^N {\sf P}(Z_i | Q_{i-1})
\prod_{i=1}^N {\sf P}(Q_i | Q_{i-1}, \hsppp Z_i ) 
\prod_{i=1}^N {\sf P}(\Delta_i | Q_i).
\label{eqn5}
\end{eqnarray}
The dependence relations that drive the coupled HMM framework for
user activity are illustrated in Figs.~1 and~2. 

\subsection*{Comparison with Related Models and Architectures}
Many extensions to the classical HMM architecture have been proposed in the
literature for inferencing problems in different settings. We now compare
the proposed coupled HMM architecture with some of these extensions and
variants. The
conditional dependencies of the involved variables corresponding to the
different architectures relative to the proposed model in this work are
summarized in Table~3. 
The number of model parameters for these
architectures with $k$ states, $\ell$ network influence structure levels, and
$m$ parameters for the observations are presented in Table~4. 

The simplest extension of the HMM architecture, an {\em autoregressive}
HMM~\cite{juang_rabiner1}, ties the evolution of the user's inter-tweet
duration to his state and his past observations. Thus, this model does
not incorporate the influence of the user's network on his activity. A
more general framework called a class of {\em input-output} HMMs is proposed
in~\cite{Cacciatore94mixturesof} and~\cite{bengio}, where the user's
inter-tweet duration is not only dependent on his state, but also on an
{\em external} input such as the network influence structure. In contrast,
a coupled variation of the {\em factorial} HMM in~\cite{ghahramani} takes
the viewpoint of the network influence structure being an additional state
rather than an external input. In either model, the current state of the user
depends not only on his past state, but also on the state of his network.
However, in both models, the evolution of the user's network is
{\em one-sided} and independent of the user's interaction.

The case of a {\em fully-coupled} HMM overcomes this one-sided evolution
by ensuring that the state of the user and his network {\em influence} each
other. The price to pay for such generality (lack of structure in the conditional
dependencies) is that the model parameters have to be learned via approximation
algorithms instead of iterative techniques. To overcome this difficulty, a
structured architecture is proposed in~\cite{brand_oliver_pentland}, where
\begin{eqnarray}
{\sf P}(Q_i | \left\{ Q_{i-1}, \hsppp Z_{i-1} \right\} ) & = &
{\sf P}(Q_i | Q_{i-1} ) \cdot {\sf P}(Q_i | Z_{i-1} ) \nonumber \\
{\sf P}(Z_i | \left\{ Q_{i-1}, \hsppp Z_{i-1} \right\} ) & = &
{\sf P}(Z_i | Q_{i-1} ) \cdot {\sf P}(Z_i | Z_{i-1} ). \nonumber
\end{eqnarray}
Similarly,~\cite{zhong_ghosh} proposes another architecture, where
\begin{eqnarray}
{\sf P}(Q_i | \left\{ Q_{i-1}, \hsppp Z_{i-1} \right\} ) & = &
\beta_1 \cdot {\sf P}(Q_i | Q_{i-1} ) + \beta_2 \cdot {\sf P}(Q_i | Z_{i-1} ) \nonumber \\
{\sf P}(Z_i | \left\{ Q_{i-1}, \hsppp Z_{i-1} \right\} ) & = &
\gamma_1 \cdot {\sf P}(Z_i | Q_{i-1} ) + \gamma_2 \cdot {\sf P}(Z_i | Z_{i-1} ) \nonumber
\end{eqnarray}
for appropriate normalization constants $\{ \beta_{i} \}$ and $\{ \gamma_{i} \}$.
While parameter learning algorithms simplify in either scenario, from a social
network perspective, the number of model parameters remain large for small values
of $k$ and $\ell$ relative to the proposed coupled HMM architecture in this work.
For example, these structured coupled HMMs are described by $10$ and $18$ model
parameters with an $m = 1$ parameter observation density in each state (and $12$
and $20$ parameters with $m = 2$) relative to $8$ and $10$ model parameters with
the proposed model in the same settings (see Table~4 
for details).

In the backdrop of this discussion, the proposed coupled HMM architecture
offers a principled, novel and easily-motivated modeling framework, specifically
useful in social network contexts. Despite being simple, it offers significant
performance gains over the classical HMM architecture and its variants. As
the subsequent discussion also shows, the proposed architecture has the
added benefit of allowing model learning via simple re-estimation formulas.

\ignore{
lists
the number of parameters that capture the different models as a function
of $k$ and $m$. Specifically\footnote{In these calculations, the initial
state probability parameters $\{ \pi_i \}$ are not counted as
model-determinants since numerical results show no dependence on the
particular initialization of $\{ \pi_i \}$.}, a simple HMM with $k = 2$
states and an $m = 1$ parameter observation density is described by $4$
parameters, whereas a coupled HMM with an additional one parameter influence
structure is described by $8$ parameters.
}

\section*{Methodology}
\label{sec4}

\subsection*{Learning Model Parameters}
\label{sec4a}
It is of interest to infer the underlying states $\{ Q_i \}$ that
cannot be observed directly. This task is performed with the aid of
the observations $\{ \Delta_i \}$ in the HMM setting, and with the aid
of $\{ \Delta_i \}$ and the influence structure $\{ Z_i \}$ in the coupled
HMM setting.

In the HMM setting, a {\em locally} optimal choice of model parameters is
sought to maximize the likelihood function ${\sf P}(\Delta_1^N |
\boldsymbol\lambda)$. Following the result in~\cite{dempster,liporace}, starting
with an initial choice of HMM parameters ${\boldsymbol{\bar{\lambda}}}$, the
model parameters are re-estimated to maximize Baum's auxiliary
function $Q(\boldsymbol\lambda, \hsppp {\boldsymbol{\bar{\lambda}}}) \Big|_
{\sf HMM}$, defined as,
\begin{equation}
\begin{split}
& {\hspace{-0.05in}}
Q(\boldsymbol\lambda, \hsppp {\boldsymbol{\bar{\lambda}}})
\Big|_ {\sf HMM} \triangleq
\sum_{ Q_0^N }
\log \Big( {\sf P}  \left( \Delta_1^N, Q_0^N | \boldsymbol\lambda
\right) \Big) \cdot {\sf P} \Big( \Delta_1^N, Q_0^N |
{\boldsymbol{\bar{\lambda}}} \Big). 
\nonumber
\end{split}
\end{equation}
It can be easily checked~\cite{rabiner,bilmes} that this maximization
breaks into a term-by-term optimization of individual model parameters.
The model re-estimation formulas are given by the Baum-Welch algorithm:
\ignore{
\begin{eqnarray}
\widehat{s}_0 & = & \frac{ \sum _{i=1}^N \xi_{i-1}(0,1) }
{ \sum _{i=1}^N \xi_{i-1}(0,0) + \sum _{i=1}^N \xi_{i-1}(0,1) },
\nonumber \\
\widehat{t}_0 & = & \frac{ \sum _{i=1}^N \xi_{i-1}(1,0) }
{ \sum _{i=1}^N \xi_{i-1}(1,0) +
\sum _{i=1}^N \xi_{i-1}(1,1) }. \nonumber 
\end{eqnarray}
}
\begin{equation}
\begin{split}
& \widehat{\beta}_{i,j} = \frac{ \sum _{i=1}^N \xi_{i-1}(i,j) }
{ \sum _{i=1}^N \left( \xi_{i-1}(i,0) + 
\xi_{i-1}(i,1) \right) },
\hspp \hspp i \neq j, \hspp \hspp i, j \in \{0,1\}.
\nonumber
\end{split}
\end{equation}
The Baum-Welch estimate of the parameters defining the different
observation densities are presented in Table~1. 
In these equations,~$\xi_i(a,b)$ for $i = 0, \cdots, N-1$ is defined as
\begin{eqnarray}
\xi_i(a,b) \triangleq {\sf P} \left(Q_{i} = a, Q_{i+1} = b |
\Delta_1^N , {\boldsymbol{\bar{\lambda}}} \right).
\nonumber
\end{eqnarray}
The update equation for $\xi_i(a,b)$ follows from~\cite[Eq.~(37)]{rabiner}
and the forward-backward procedure.

In the coupled HMM setting, as in~\cite{bengio}, we are interested in model
parameters that maximize the {\em conditional} likelihood function
${\sf P}(\Delta_1^N |Z_1^N , \hsppp \boldsymbol\lambda)$. If the conditional
likelihood is known in closed-form (as in the input-output HMM case~\cite{bengio})
or a tight lower bound to it is known, a conditional expectation maximization
algorithm along the lines of~\cite{jordan_jacobs,jebara_pentland,salojarvi}
can be pursued. In the proposed coupled HMM setting, the conditional likelihood
appears to be neither amenable to a simple formula nor a tight lower bound. To
overcome this technical difficulty, we now propose a two-step procedure to learn
an estimate of the model parameters that maximize ${\sf P}(\Delta_1^N |
Z_1^N , \hsppp \boldsymbol\lambda)$.

In the first step, we fix the threshold that distinguishes the baseline dynamics
${\sf P}_{\sf 0}$ from the elevated dynamics ${\sf P}_{\sf 1}$, $\tau$, to an
appropriate choice $\boldsymbol\tau_{\sf init}$. We then treat $\Delta_1^N$
and $Z_1^N$ as training observations and consider a generalized auxiliary
function $Q(\boldsymbol\lambda, \hsppp {\boldsymbol{\bar{\lambda}}})
\Big|_{\sf CHMM}$ of the form:
\begin{equation}
Q(\boldsymbol\lambda, \hsppp
{\boldsymbol{\bar{\lambda}}}) \Big|_{\sf CHMM} \triangleq
\sum_{ Q_0^N } \log \Big( {\sf P}  \left(
\Delta_1^N, Z_1^N, Q_0^N | \boldsymbol\lambda  \right)
\Big) \cdot {\sf P} \Big( \Delta_1^N, Z_1^N, Q_0^N |
{\boldsymbol{\bar{\lambda}}} \Big).
\label{eqn6}
\end{equation}
As before, $\boldsymbol\lambda$ and ${\boldsymbol{\bar{\lambda}}}$ denote the
optimization variable corresponding to the parameter space (all the parameters
except for $\tau$) and its initial estimate, respectively. A straightforward
extension of the proof in~\cite{dempster,liporace} shows that maximizing
$Q(\boldsymbol\lambda, \hsppp {\boldsymbol{\bar{\lambda}}}) \Big|_{\sf CHMM}$
in the $\boldsymbol\lambda$ variable results in a local maximization of the
{\em joint} likelihood function ${\sf P} \left( \Delta_1^N, Z_1^N | \boldsymbol
\lambda \right)$. To obtain analogous re-estimation formulas for an iterative
solution to a local maximum, we define the equivalent intermediate variable $\widetilde{\xi}_i(a,b)$ for $i = 0, \cdots, N-1$:
\begin{eqnarray}
\widetilde{\xi}_i(a,b) \triangleq
{\sf P} \left(Q_{i} = a, \hsppp Q_{i+1} = b | \Delta_1^N , Z_1^N,
{\boldsymbol{\bar{\lambda}}} \right). \nonumber
\end{eqnarray}
Using~(\ref{eqn5}), we can simplify~(\ref{eqn6}) and the joint optimization
of the model parameters again breaks into a term-by-term optimization.
We then have the following analogous model parameter estimates for
$k \in \{0 , 1\}$:
\begin{equation}
\begin{split}
& {\hspace{0.05in}}
\widetilde{p}_k = \frac{ \sum _{ i = 1, \hspp i \hsppp \in \hsppp {\cal Z}_k}^N
\widetilde{\xi}_{i-1}(0,1) }
{ \sum _{ i = 1, \hspp i \hsppp \in \hsppp {\cal Z}_k}^N \widetilde{\xi}_{i-1}(0,0)
+ \sum _{ i = 1, \hspp i \hsppp \in \hsppp {\cal Z}_k}^N \widetilde{\xi}_{i-1}(0,1) }
\\ & {\hspace{0.05in}}
\widetilde{q}_k =
\frac{ \sum _{ i = 1, \hspp i \hsppp \in \hsppp {\cal Z}_k}^N
\widetilde{\xi}_{i-1}(1,0) }
{ \sum _{ i = 1, \hspp i \hsppp \in \hsppp {\cal Z}_k}^N \widetilde{\xi}_{i-1}(1,0)
+ \sum _{ i = 1, \hspp i \hsppp \in \hsppp {\cal Z}_k}^N \widetilde{\xi}_{i-1}(1,1) }
\nonumber
\end{split}
\end{equation}
with ${\cal Z}_0 = \{ i \hsppp : \hsppp Z_i \leq \tau \}$ and
${\cal Z}_1 = \{ i \hsppp : \hsppp Z_i > \tau \}$. The re-estimation formulas
for the observation density parameters follow the same structure as in
Table~1 
by replacing ${\xi}_{i}(\cdot,\cdot)$ with
$\widetilde{\xi}_{i}(\cdot,\cdot)$. For the parameters defining the density
of the influence structure, Table~2 
provides a list of re-estimation formulas. The intermediate
variable $\widetilde{\xi}_i(a,b), \hspp 1 \leq i \leq N-1$ is updated by a
generalized forward-backward procedure whose steps are illustrated
in Table~5 (see~(\ref{ref1_eqn_begin})-(\ref{ref1_eqn_end})).

We denote by $\boldsymbol\lambda_{\sf init}$ the converged model parameters
that locally maximize $Q(\boldsymbol\lambda, \hsppp {\boldsymbol{\bar{\lambda}}})
\Big|_{\sf CHMM}$. Note that $\boldsymbol\lambda_{\sf init}$ is a local
maximum only in the $\boldsymbol\lambda$ space and not in
$\left\{ \boldsymbol\tau \times \boldsymbol\lambda \right\}$. Thus,
the choice $\left\{ \boldsymbol\tau_{\sf init}, \hsppp \boldsymbol
\lambda_{\sf init} \right\}$ does not maximize ${\sf P}(\Delta_1^N|Z_1^N,
\boldsymbol \lambda)$, not even locally. Therefore, in the next step, we
locally optimize the conditional likelihood over a local region around
$\left\{ \boldsymbol\tau_{\sf init}, \hsppp \boldsymbol\lambda_{\sf init}
\right\}$. That is,
\begin{eqnarray}
\left\{ \widetilde{\boldsymbol\tau}, \hsppp \widetilde{\boldsymbol\lambda}
\right\}
= \argmax_{ \left\{\boldsymbol\tau, \hsppp \boldsymbol\lambda \right\}
\hsppp \in \hsppp {\cal L}} {\sf P}
\left(\Delta_1^N |Z_1^N , \hsppp \boldsymbol\lambda \right)
\nonumber
\end{eqnarray}
where ${\cal L} = \left\{ \boldsymbol\tau \hsppp : \hsppp \boldsymbol\tau
= \boldsymbol\tau_{\sf init} + \Delta \boldsymbol\tau \hspp
{\rm and} \hspp
\boldsymbol\lambda \hsppp : \hsppp \boldsymbol\lambda
= \boldsymbol\lambda_{\sf init} + \Delta \boldsymbol\lambda \right\}$. In this
work, we focus on a box-constrained ${\cal L}$. Alternately, a local gradient
search in the model parameter space can be pursued to locally optimize the
conditional likelihood function. Note that the conditional density can be
written as
\begin{eqnarray}
{\sf P} \left(\Delta_1^N |Z_1^N , \hsppp \boldsymbol\lambda \right) =
\frac{ \widetilde{\alpha}_N(0) + \widetilde{\alpha}_N(1) }
{ \widehat{\alpha}_N(0) + \widehat{\alpha}_N(1)
},  \nonumber
\end{eqnarray}
where the forward algorithm variable $\widetilde{\alpha}_i(j)$ is updated
with the same formulas as~(\ref{eq_alpha1})-(\ref{eq_alpha3}) (see Table~5).
On the other hand, $\widehat{\alpha}_i(j)$ follows the same formula as
$\widetilde{\alpha}_i(j)$, but by constraining ${\sf P}(\Delta_i|Q_i = a) = 1$
for all $i$ and $a$.

\subsection*{Model Verification}
The efficacy of the different models to the observed data are studied in
two ways.
In the first approach, the model parameters learned via the (generalized)
Baum-Welch algorithm are used with a state estimation procedure to estimate
the most probable state sequence associated with the observations. For the
HMM setting, state estimation is straightforward via the use of the Viterbi algorithm~\cite{rabiner}. The generalization of the Viterbi
algorithm to the coupled HMM setting requires the definition of intermediate
variables $\delta_{i+1}(j)$ and $\phi_{i+1}(j)$ as illustrated in Table~5~(see~(\ref{ref2_eqn_begin})-(\ref{ref2_eqn_end})). As with
the Viterbi algorithm, the most probable state sequence is then estimated as
\begin{eqnarray}
Q^{\star}_N & = & \argmax_{ j \hsppp \in \hsppp \{ 0, 1\}} \delta_N(j) \nonumber \\
Q^{\star}_i & = & \phi_{i+1}(Q^{\star}_{i+1}), \hspp 1 \leq i \leq N -1.
\nonumber
\end{eqnarray}
The observed inter-tweet durations corresponding to the classified states
are compared with the inter-tweet durations obtained with the proposed model(s)
via a graphical method such as the Quantile-Quantile (Q-Q) plot. Recall that a
Q-Q plot plots the quantiles corresponding to the true observations with
the quantiles corresponding to the model(s)~\cite{gnanadesikan}. If the
proposed model reflects the observations correctly, the quantiles lie on
the (reference) straight-line that extrapolates the first and the third quartiles.
Discrepancies from the straight-line benchmark indicate artifacts introduced by
the model(s) not seen in the observations and/or features in the observation
not explained by the model(s).

In the second approach, the fits of the different models to the data are
studied via a more formal metric such as the Akaike
Information Criterion (AIC), defined as 
\begin{eqnarray}
{\tt AIC}(n) & \triangleq & 2k - 2 \log({\sf L}), \nonumber
\end{eqnarray}
where $k$ denotes the number of parameters used in the model, $n$ is the
length of the observation sequence, and ${\sf L}$ is the optimized likelihood
function for the observation sequence corresponding to the learned model.
The AIC penalizes models with more parameters and the model that
results in the smallest value of AIC 
is the most suitable model (for the observed data) from the class
of models considered. 
In the HMM setting with $k_{\sf HMM}$ parameters, the AIC corresponding
to $\{ \Delta_i \}$ is given as
\begin{eqnarray}
{\tt AIC}(n) \Big|_{\sf HMM} & \triangleq &
2 k_{\sf HMM} - 2 \log \big(  {\sf P}(\Delta_1^N|
\boldsymbol\lambda)  \big) \nonumber \\
& = & 2 k_{\sf HMM} 
- 2 \log \Big(
\alpha_N(0) + \alpha_N(1) \Big),  \nonumber
\end{eqnarray}
where the converged model parameter estimates from the Baum-Welch
algorithm are used to compute $\alpha_i(j) \triangleq
{\sf P}(\Delta_1^i, \hsppp Q_i = j)$ using the forward procedure. In
the coupled HMM setting with $k_{\sf CHMM}$ parameters, the corresponding
AIC metric is defined as
\begin{equation}
\begin{split}
{\tt AIC}(n) \Big|_{\sf CHMM} \triangleq 2 k_{\sf CHMM} 
- 2 \log \big( {\sf P}(\Delta_1^N|Z_1^N,
\boldsymbol\lambda) \big), 
\label{bic_chmm}
\end{split}
\end{equation}
where the model parameter estimates that maximize
${\sf P}(\Delta_1^N|Z_1^N, \boldsymbol\lambda)$ are to be used
in~(\ref{bic_chmm}). 
With the model parameters learned as explained in the previous section,
an upper bound to ${\tt AIC}(n) \Big|_{\sf CHMM}$ is obtained
as 
\begin{eqnarray}
{\tt AIC}(n) \Big|_{\sf CHMM} \leq {\overline{ \tt AIC}}(n)
\triangleq 2 k_{\sf CHMM} - 2 \log \left(
\frac{ \widetilde{\alpha}_N(0) + \widetilde{\alpha}_N(1) }
{ \widehat{\alpha}_N(0) + \widehat{\alpha}_N(1)
} \right). \nonumber
\end{eqnarray}

\subsection*{Forecasting}
Given $\Delta_1^n$ (and $Z_1^n$), forecasting $\Delta_{n+1}$ is of immense
importance in tasks such as advertising, anomaly detection (detecting when a
compromised account will post next), etc. A simple maximum {\em a posteriori}
(MAP) predictor of the form
\begin{eqnarray}
\widetilde{\Delta}_{n+1} \Big|_{\sf MAP} 
= \argmax_{y} f(\Delta_{n+1} = y | \Delta_1^n, Z_1^n )
= \argmax_{y} \sum_{i = 0}^{k-1}
\widetilde{\beta}_i f_i( \Delta_{n+1} = y  ) 
\nonumber
\end{eqnarray}
where
\begin{eqnarray}
\widetilde{\beta}_i = \frac{ \sum_j \widetilde{\alpha}_n(j)
{\sf P}( {Z}_{n+1} | Q_n = j) {\sf P}(Q_{n+1} = i |
{Z}_{n+1}, Q_n = j) } { \sum_j \widetilde{\alpha}_n(j) }
\nonumber
\end{eqnarray}
fails when $f_i(\Delta_{n+1} = y)$ is unimodal with the same mode for all $i$.
This is always the case with exponential observation models (mode is $0$)
and with gamma models if $k_i \lambda_i < 1$ for all $i$ (mode is $0$), which
is typically the case with the best model fits for many users. On the other
hand, a conditional mean predictor of the form
\begin{eqnarray}
\widetilde{\Delta}_{n+1} \Big|_{\sf CM} ={\sf E}
\left[ \Delta_{n+1} | \Delta_1^n, Z_1^n \right] =
\sum_{i = 0}^{k - 1} \widetilde{\beta}_i {\sf E} \left[ \Delta_{n+1}
| Q_{n+1} = i \right] \nonumber
\end{eqnarray}
results in large forecasting errors in the {\em Inactive} state if the
mean inter-tweet durations in the two states are very disparate (typically
the case for many users). To overcome these problems, we consider a predictor
of the form
\begin{eqnarray}
\widetilde{\Delta}_{n+1}  = \sum_{i = 0}^{k - 1}
\indic \left( \widetilde{Q}_{n+1} = i|\Delta_1^n \right)
{\sf E} \left[ \Delta_{n+1} | Q_{n+1} = i \right]
\nonumber
\end{eqnarray}
where $\widetilde{Q}_{n+1}$ is the state estimate using the (generalized)
Viterbi algorithm with $\Delta_1^n$ (and $Z_1^n$) as inputs and
study the forecasting performance in the {\em Active} state with a
Symmetric Mean Absolute Percentage Error (SMAPE) metric:
\begin{eqnarray}
{\tt SMAPE}(N) \triangleq \frac{1}{N} \sum_{i = 1}^N
\left|
\frac{ \Delta_i - \widetilde{\Delta}_i } {\Delta_i + \widetilde{\Delta}_i }
\right| \cdot \indic(Q_i = 1).
\nonumber
\end{eqnarray}
The SMAPE metric is a normalized error metric and a smaller value
indicates a better model for forecasting. It is seen as a percentage
error and is bounded between $0 \%$ and $100 \%$. 

\section*{Numerical Results}
\label{sec5}
The dataset used to illustrate the efficacy of the models proposed in this work
is a $30$-day long record of ${\tt Twitter}$ activity described in~\cite{sofus}. This
dataset consists of ${\cal N}_{\sf t} = 652,522$ tweets from ${\cal N}_{\sf u}
= 30,750$ users (with at least one tweet). The time-scale on which the tweets
are collected is minutes. While the dataset has been collected using a
snowball sampling technique and reflects a population primarily based out of
West Asia, London and Pakistan, the users in the dataset appear to be from
diverse socio-economic and political backgrounds and have a broad array of
interests. Further, the properties of the dataset on a collective scale are
similar in nature to well-understood properties of similar datasets~\cite{sofus}
suggesting a high confidence on the suitability of the dataset in studying
user behavior on an individual scale and in its generalizability to other datasets.

Since reliable model learning can be accomplished only for users with sufficient
activity, we focus on users with a large number of tweets over the data collection
period. There were $223$ users with over $600$ tweets and $115$ users with over
$1,000$ tweets.

\subsection*{Validating Model Assumptions}
The coupled HMM framework developed in this paper is built on four main
assumptions, two of which lead to the HMM formulation and two that couple the
influence of the neighbors to the HMM. Notwithstanding the fact that these
assumptions are based on a rational model of user behavior, the first two
assumptions have been well-studied and justified in the
literature~\cite{malmgren,wu_zhou,hogg_lerman,Romero2010}. We now provide
some empirical results to justify the latter two assumptions.

For this, 
we start with two typical users (denoted as User-I and User-II) whose
activity over the thirty-day period consists of: i) $807$ tweets, $260$
mentions, and $16,935$ tweets from his social network of $62$ friends, and
ii) $1,914$ tweets, $1,108$ mentions, and $10,281$ tweets from his social
network of $92$ friends.
We also consider an extreme case of a highly active user (denoted as User-III)
whose activity over the thirty-day period consists of $2,387$ tweets, $2,872$
mentions, and $58,810$ tweets from his social network of $206$ friends. Users-I
and II do not appear to be popular public figures, whereas User-III is a popular
journalist, advocate on many political issues, and an activist.

Assumption~3 hypothesizes that each user's activity switches from a baseline dynamics
to an elevated state of dynamics depending on the magnitude of $\tau$. To test
this assumption, we study the data from User-III for $n = 1000$. With this data,
we use the generalized Baum-Welch algorithm to learn model parameters (as a
function of $\tau$) for a coupled HMM with the number of mentions as the influence
structure. Fig.~3 
plots the learned transition probabilities for User-III as a function of $\tau$.
From this figure, we see that both $p_0$ and $p_1$ start off around approximately
the same value (and similarly for $q_0$ and $q_1$). However, as $\tau$ increases,
the transition probabilities stabilize at different values suggesting that there
is indeed a baseline and an elevated state in user dynamics. Similar behavior is
also seen with data from Users-I and II.

On the other hand, Assumption~4 hypothesizes that $Z_i$ and $Z_{i-1}$ are
conditionally independent given $Q_{i-1}$. To test this assumption, we use the
model parameters learned with the generalized Baum-Welch algorithm in the
generalized Viterbi algorithm to estimate the most likely state sequence
corresponding to the observations. The conditional correlation coefficient
between $Z_i$ and $Z_{i-1}$, defined as,
\begin{eqnarray}
\rho (Z_i , \hsppp Z_{i-1} | Q_{i-1} = j) \triangleq
\frac{ {\sf E} \Big[ \big( Z_i - {\sf E}[Z_i | Q_{i-1} = j] \big)
\cdot
\big( Z_{i-1} - {\sf E}[Z_{i-1} | Q_{i-1} = j] \big) \Big]  }
{ \sqrt{ {\sf E} \left[ \left( Z_i - {\sf E}[Z_i | Q_{i-1} = j] \right)^2 \right]}
\cdot
\sqrt{ {\sf E} \left[ \left( Z_{i-1} - {\sf E}[Z_{i-1} | Q_{i-1} = j] \right)^2 \right]}
} , \hspp j \in \{ 0, 1 \}
\nonumber
\end{eqnarray}
is used to study conditional independence. Table~6 
lists $\rho (Z_i , \hsppp Z_{i-1} | Q_{i-1} = j)$ and the $p$-value corresponding
to this coefficient for Users-I to III with $n = 800$, $n = 1000$ and $n = 1000$,
respectively. From this table, we see that the correlation coefficient in all the
six cases studied has a small (absolute) value. A simple explanation for this
observation is that user aggregation significantly diminishes the correlation
between $Z_{i-1}$ and $Z_i$. Specifically, at a (standard) significance level of
$5 \%$, the null hypothesis of conditional independence between $Z_i$ and
$Z_{i-1}$ cannot be rejected in five of the six cases studied indicating that
Assumption~4 can be justified for many users.

\subsection*{Model Fits For Users-I to III}
We now study the following models for the activity profile of the three users: i)
conventional two-state HMM, ii) coupled HMM with a binary influence structure
that is set to $1$ when there is a mention of the user and $0$ otherwise, iii)
coupled HMM with the number of such mentions as the influence structure, and
iv) coupled HMM with the social network traffic of the friends of the
user as the influence structure. Exponential and gamma densities are
considered for the observations (inter-tweet duration). On the other hand,
geometric, Poisson and shifted zeta densities are considered for the number of
mentions, and a geometric density is considered for the total traffic. See
Tables~1 
and~2 
for model details.

Tables~7-9 
list the AIC scores for these three users with
the different models as a function of the number of observations $n$ for
different choices: $n = 100$, $n = 250$, $n = 500$, $n = 750$, or $n = 1000$.
From these tables, the following conclusions can be made:
\begin{enumerate}
\item
For all the three users, both a Poisson process model and a renewal process
model are {\em significantly} sub-optimal for the observations as they
implicitly assume a single state for the user's activity. This is in conformance
with similar observations in~\cite{malmgren}. A conventional HMM with two
states overcomes this problem by assuming that the user switches between an
{\em Active} and an {\em Inactive} state and thus provides a better baseline to
compare the performance of the proposed modeling framework.

\item
For all combinations of users, $n$ and types of influence structure, a
two-parameter gamma density for the observations results in a better fit
than possible with an exponential model. This should not be entirely
surprising since an exponential density is a special case of the gamma
density (a gamma with $k = 1$ and $\lambda = \frac{1}{\rho}$ results in
an exponential of rate $\rho$). Similar observations have also been made
in related recent work~\cite{xiao_guo_icdcs2007,guo_tan_podc2008,guo_tan_kdd2009,jiang}.

This conclusion is also reinforced by observing the Q-Q plots of the true
inter-tweet durations (in the {\em Active} and {\em Inactive} states)
relative to the inter-tweet duration values obtained from four models
for User-III (see Figs.~4(a)-(d) 
and Figs.~4(e)-(h), 
respectively). The four competing models illustrated are: i) Model
{\sf A} --- conventional HMM with exponential density, ii) Model
{\sf B} --- conventional HMM with gamma density, iii) Model
{\sf C} --- coupled HMM with geometric influence structure and exponential
density, and iv) Model {\sf D} --- coupled HMM with geometric influence
structure and gamma density. The most probable state sequence vector
corresponding to the observations of User-III is estimated with the
(generalized) Viterbi algorithm. As can be seen from Fig.~4, 
Model {\sf D} is the best fit from among these four models. Nevertheless,
the discrepancies of some of the quantiles from the reference straight-line
shows that even this model does not {\em completely} capture all the features
in the observations, suggesting a direction for future work in this area.

\item
We now explain how a coupled HMM works. State estimation with Models {\sf B}
and {\sf D} is performed using the (generalized) Viterbi algorithm. In the
$n = 1000$ case, while the HMM declares $898$ of the $1000$ inter-tweet periods
(corresponding to $21.06 \%$ of the total observation period for User-III)
as {\em Active}, the coupled HMM declares only $845$ periods (corresponding
to $16.46 \%$ of the total observation period) as {\em Active}. In terms of
discrepancies in state estimation between the two models, $53$ inter-tweet
periods declared as {\em Active} by the HMM are re-classified as {\em Inactive}
by the coupled HMM, whereas all the {\em Inactive} states of the HMM are also
classified as {\em Inactive} by the coupled HMM.
Carefully studying $\Delta_i$, $Q_{i-1}$ and $Z_i$ for these $53$ periods that
are re-classified, it can be seen that the coupled HMM declares a period as
{\em Inactive} (independent of the nature of $Q_{i-1}$ or $Z_i$) provided that
$\Delta_i$ is large, or when $\Delta_i$ is small and in addition, $Z_i$ is also
small and $Q_{i-1} = 0$. In other words, if the user is in the {\em Inactive}
state and the influence structure does not suggest a switch to the {\em Active}
state, a small inter-tweet period is treated as an {\em anomaly} rather than as
an indicator of change to the {\em Active} state. Thus, in contrast to the HMM
setting where the state estimate depends primarily on the magnitude of $\Delta_i$,
the coupled HMM is less {\em trigger-happy} in the sense that it considers the
magnitude of $\Delta_i$ in the context of neighbors' activity
before declaring a state as {\em Active} or {\em Inactive}.

\item
In terms of general trends with AIC as the metric for model fitting, if $n$ is
small (say, $n = 100$ to $250$), a conventional HMM is competitive and comparable
with (sometimes, even better than) a coupled HMM with more parameters. In
addition to there not being sufficient data to learn a complicated model, this
trend conforms with the popular intuition of the Occam's razor that simplistic
models shall suffice for observations of small length.

\item
Social network traffic (that includes replies, retweets, and modified tweets
from all of a user's friends) typically overwhelms the number
of mentions by at least an order of magnitude (see typical examples with
Users-I and II above). Thus, social network traffic serves as a good influence
structure to couple a HMM when the number of mentions is too small to learn a
sophisticated model reliably. This is typically the case when $n$ is moderate
(neither too small nor too large). For example, with Users-I and II, traffic
leads to a better model fit for $n = 500$ and $n = 100$, respectively.

\item
However, as $n$ increases, the more {\em directional} nature of a mention
(relative to the traffic) means that mentions carry more ``information''
about the capacity of a user to respond/reply conditioned on seeing a certain type
of tweet from his network (than the traffic). This is clear from the general
trend of lower AIC scores with the number of mentions than with the social
network traffic for large $n$ values ($n = 750$ or $1000$).

\item
For all combinations of users and $n$, the binary influence structure for the
mentions results in a poorer fit than the number of mentions. This is because
it is more efficient to capture the number of mentions with a one parameter
model than to expend that parameter on a binary value. In other words, the loss
in performance is due to the use of a hard decision metric (binary value),
provided that the soft decision metric (the number of mentions) is captured
accurately. While the three models (geometric, Poisson and shifted zeta)
for the number of mentions result in comparable performance for Users-I and
II, the geometric results in a superior fit for User-III. Thus, a geometric
density can serve as a robust model choice for the number of mentions. Given
that the shifted zeta density captures heavy-tails, the above trend also
suggests that the number of mentions over an inter-tweet duration is not likely
to be heavy-tailed.
\end{enumerate}

\subsection*{Performance Across Users}
Given that the exponential observation density consistently under-performs
in model fitting relative to a gamma density, we henceforth focus on the
performance of Model {\sf D} with Model {\sf B} as the baseline. This
performance gain is captured by the relative AIC gain metric, defined as,
\begin{eqnarray}
\Delta {\tt AIC} & \triangleq &
{\tt AIC} \Big|_{ {\sf Model \hspp B} } -
{\tt AIC} \Big|_{ {\sf Model \hspp D} }.
\label{delta_aic}
\end{eqnarray}
In Fig.~5 and Table~10, 
$\Delta {\tt AIC}$ values are presented for different $n$ values for
Users-I to III. As can be seen from this data, Model {\sf B} performs
better than Model {\sf D} for $n < 550$ for User-I and Model {\sf D}
gets better as $n$ increases after that. For User-III, Model {\sf D}
is better than Model {\sf B} for all $n > 200$ and the performance
gain improves with increasing $n$ till $n = 800$ and then slightly
decreases after that. In general, the performance gain with Model {\sf D}
for the typical user is negligible for small values of $n$ and this gain
improves (in general) as $n$ increases.

To study this aspect more carefully, we now consider a corpus of $100$ users
with different numbers of tweets and mentions over their periods of activity.
For all the users studied, it is observed that a local optimum (to reasonable
accuracy) is achieved by the generalized Baum-Welch algorithm within $20$-$30$
iterations and independent of the model parameter initializations.
Fig.~6(a)-(b) 
plots the histogram of $\Delta {\tt AIC}$ for the corpus of $100$ users with
$n = 500$ and $n = 1000$, respectively. From Fig.~6, 
it can be seen that Model {\sf D} significantly out-performs Model {\sf B} for a
large fraction of the users and this improvement gets better as $n$ increases.

To understand this, recall that $\exp \left( - \frac{ \Delta {\tt AIC} }{2}
\right)$ is the likelihood that Model {\sf B} minimizes the information loss
relative to Model {\sf D}. Thus, a $\Delta {\tt AIC}$ value larger than $4.61$
and $9.21$ leads to a relative likelihood of $10 \%$ and $1 \%$, respectively.
For the corpus studied here, Model {\sf D} is $100$ times as likely to minimize
information loss for $25 \%$ of the users at $n = 500$ and $72 \%$ of the users
at $n = 1000$, respectively. With a more relaxed benchmark, Model {\sf D} is ten
times as likely to minimize information loss for $33 \%$ and $85 \%$ of the users
at $n = 500$ and $1000$, respectively.

For predictive performance, analogous to $\Delta {\tt AIC}$ in~(\ref{delta_aic}),
we define the relative SMAPE gain metric as
\begin{eqnarray}
\Delta {\tt SMAPE} & \triangleq &
{\tt SMAPE} \Big|_{ {\sf Model \hspp B} } -
{\tt SMAPE} \Big|_{ {\sf Model \hspp D} }.
\nonumber
\end{eqnarray}
Fig.~6(c) 
plots the histogram of $\Delta {\tt SMAPE}$ for $n = 500$ and it can again be
seen that Model {\sf D} is better than Model {\sf B} in terms of predictive
power for a large fraction of users. Thus, a coupled HMM provides a better modeling
paradigm for the activity of a large set of users in social networks.

\ignore{
In general, the following conclusions can be made based on our
studies: i) Model d would be most
useful if there are enough observations and influence structure
observations 
to ensure the accurate learning of the sophisticated model, ii) Model b
would be most useful if there are enough observations, but not enough
influence structure observations, 
iii) The simplest choice, Model a, would be most useful for very limited observations.
}

\subsection*{User Clustering}
After learning the model parameters for each user, we now consider the similarity
between different users as implied by those parameters. Since the dataset
from~\cite{sofus} has only around $200$ users with sufficient activity ($n > 500$)
to learn general probabilistic models where the coupled HMM parameters learned via
the generalized Baum-Welch algorithm converge to local optima in the model
parameter space, we focus on a corpus of $150$ of these users. The mean number of
tweets and mentions for this corpus is $975.40$ and $629.17$, respectively. The
mean number of friends and followers for the corpus is $251.83$ and $206.46$,
respectively.

With the number of mentions as the influence structure, we focus on two coupled
HMM-specific parameters 
that capture the interaction dynamic between a user and his social network to
perform user clustering in the model parameter space. The parameter $p_1/p_0$
measures the propensity (likelihood) of a user to become active upon seeing a
large number of mentions in his timeline relative to a lack of such mentions:
\begin{eqnarray}
\frac{p_1}{p_0} = \frac{ {\sf P}(Q_i = 1 | Q_{i-1} = 0, \hsppp Z_i > \tau) }
{ {\sf P}(Q_i = 1 | Q_{i-1} = 0, \hsppp Z_i \leq \tau) } .
\nonumber
\end{eqnarray}
On the other hand, the parameter $\gamma_1/\gamma_0$ measures the propensity of
the user's network to respond with a mention upon seeing activity at the user
relative to his inactivity:
\begin{eqnarray}
\frac{\gamma_1}{\gamma_0} = \frac{ {\sf P}(Z_i > 0 | Q_{i-1} = 1 )}
{ {\sf P}(Z_i > 0 | Q_{i-1} = 0 )}. \nonumber
\end{eqnarray}

Three natural clusters can be identified in the model parameter space as a
function of the $\left\{ \frac{p_1}{p_0} , \hsppp  \frac{\gamma_1}{\gamma_0}
\right\}$ values: 1) The baseline scenario where the users are {\em not}
significantly influenced by their neighbors and {\em vice versa} corresponds
to $\frac{p_1}{p_0} = 1 = \frac{\gamma_1}{\gamma_0}$. 
The users for whom $\frac{p_1}{p_0} < 1$ are the ``tails'' in the model space
corresponding to
this baseline scenario. 2) On the other hand, a large value of $\frac{p_1}{p_0}$
indicates that more mentions can induce a user to the {\em Active} state.
Restated, a user can be induced to post at a higher frequency by an active
social network. 3) Similarly, a large value of $\frac{\gamma_1}{\gamma_0}$
indicates that a user's social network can be induced to become active (with a
larger number of mentions) by the user's activity. Motivated by this argument
of three natural clusters, Fig.~7 
clusters $150$ users using the $K$-means algorithm for
$K = 3$. The result of this clustering is that $92$ users belong to Cluster 1
centered around $\left( \frac{p_1}{p_0}, \hsppp \frac{\gamma_1}{\gamma_0} \right)
= (1, \hsppp 1)$. Of the remaining $58$ users, $35$ belong to Cluster 2 where
$\frac{p_1}{p_0} > 1.5$ and $23$ belong to Cluster 3 where $\frac{\gamma_1}{\gamma_0}
> 1.5$.

To paraphrase the above discussion, Cluster 1 is made of a majority of the users
corresponding to the baseline scenario. On the other hand, Clusters 2 and 3 consist
of users outside Cluster 1 and those who are either tightly knit to their network
(or {\em vice versa}). These users are significantly affected by their social
influence. Some of the typical attributes/qualities that can best describe users
in Cluster 2 are: commentarial, activist, garrulous, argumentative, opinionated,
etc. Illustrating these facets, a sample activity listing over a single session
of a typical user in Cluster 2 is provided in Table~11. 
The session begins with a question of ``yeh.watching'' (apparently, a cricket match between
Pakistan and England) by a friend of the user. This is followed by a conversation
between friends and unsolicited commentaries/observations on the ongoing match by the
user-of-interest. Such active commentary is typical of this user's posting
behavior. Another sample argument between two users in Cluster 2 is provided
in Table~12. 
This argument is an exchange of political opinions with each user trying to
convince the other about their respective positions. At the end of the argument,
one of the users realizes and acknowledges that he has become more vocal on social
media, yet also sensitive to other users' positions.

On the other hand, the social network of users in Cluster 3 share similar
attributes as users in Cluster 2 even though the users themselves are often
more reluctant to follow suit. To illustrate this subtle difference in
behaviors, a sample activity listing of a typical user in Cluster 3 is presented
in Table~13. 
Here, we see that the user's social network
is strongly opinionated in response to a news story introduced by the user.
Despite introducing the story, the user himself is not sufficiently
polarized/aggressive in his response on social media. Similarly, after
introducing another story, the user blindly agrees with other users' positions
and jokes on the matter. Thus, these examples illustrate how the coupled HMM
paradigm introduced in this work captures broad features on user behavior
despite not capturing the textual content in any detail.

\section*{Conclusion}
\label{sec6}
We have introduced a new class of coupled Hidden Markov Models to describe
temporal patterns of user activity which incorporate the social effects of
influence from the activity of a user's neighbors.
While there have been many works on models for user activity in diverse social
network settings, our work is the first to incorporate social network influence
on a user's activity. We have shown that the proposed model results in
better explanatory and predictive power over existing baseline models such as
a renewal process-based model or an uncoupled HMM. User clustering in the model
parameter space resulted in clusters with distinct interaction dynamics between
users and their networks. Specifically, three clusters corresponding to: a
baseline scenario of no influence of a user on his network (and {\em vice versa}),
and two clusters with significant influence of a user on his network, and the
network on the user, respectively are identified.

While our work has developed a social network-driven user activity model,
it has only scratched the surface in this promising arena of research. It
would be useful to pursue a more detailed study of different candidate
models for the influence structures and the observations. It is also of
interest in understanding which type of tweets/posts (mentions, replies,
retweets, undirected tweets) or the total traffic carries more ``information''
in developing good models for explanation and prediction at the individual
scale. It would also be of interest to develop hierarchical social
influence-driven models for groups of users as well as better understand
those facets of a user's social network that influence him the most. In
particular, a careful study of other network influence structures (such as
transfer entropy-weighted traffic~\cite{VerSteeg2012WWW}) that can capture
the interaction dynamic between a user and his network is of importance.

Combining temporal activity patterns with unstructured information such as
the topic or nature of discussion, textual content, etc., could result in
much better predictive performance than temporal activity alone. Further,
understanding the contribution of the temporal and the textual parts of such
a composite model in prediction would also be useful in understanding the
limits and capabilities of activity profile modeling at the individual scale.

\bigskip

\section*{List of Abbreviations}
\begin{enumerate}
\item HMM -- Hidden Markov Model
\item Coupled HMM -- Coupled Hidden Markov Model
\item Q-Q Plot -- Quantile-Quantile Plot
\item AIC -- Akaike Information Criterion
\item SMAPE -- Symmetric Mean Absolute Percentage Error
\end{enumerate}

\section*{Competing Interests}
The authors declare that they have no competing interests.

\section*{Authors' Contributions}
VR performed the numerical experiments. All the authors developed the model,
designed the experiments, interpreted the results, wrote the paper, and approved
the final manuscript.

\section*{Acknowledgment}
This work was supported by the Defense Advanced Research Projects Agency (DARPA)
under grant \# W911NF-12-1-0034 at the University of Southern California.
The authors would like to thank Sofus Macskassy for providing the
${\tt Twitter}$ dataset that was used in illustrating the algorithms proposed
in this work.



{\ifthenelse{\boolean{publ}}{\footnotesize}{\small}
\bibliographystyle{bmc_article}
\bibliography{coupled_hmm1_with_figs.bbl}}


\ifthenelse{\boolean{publ}}{\end{multicols}}{}



\newpage
\section*{Figures}

\subsection*{Figure 1 - Coupled HMM}
Coupled HMM framework for user activity.

\begin{figure}[htb!]
\centering
\begin{tabular}{c}
\includegraphics[height=2.2in,width=5.0in] {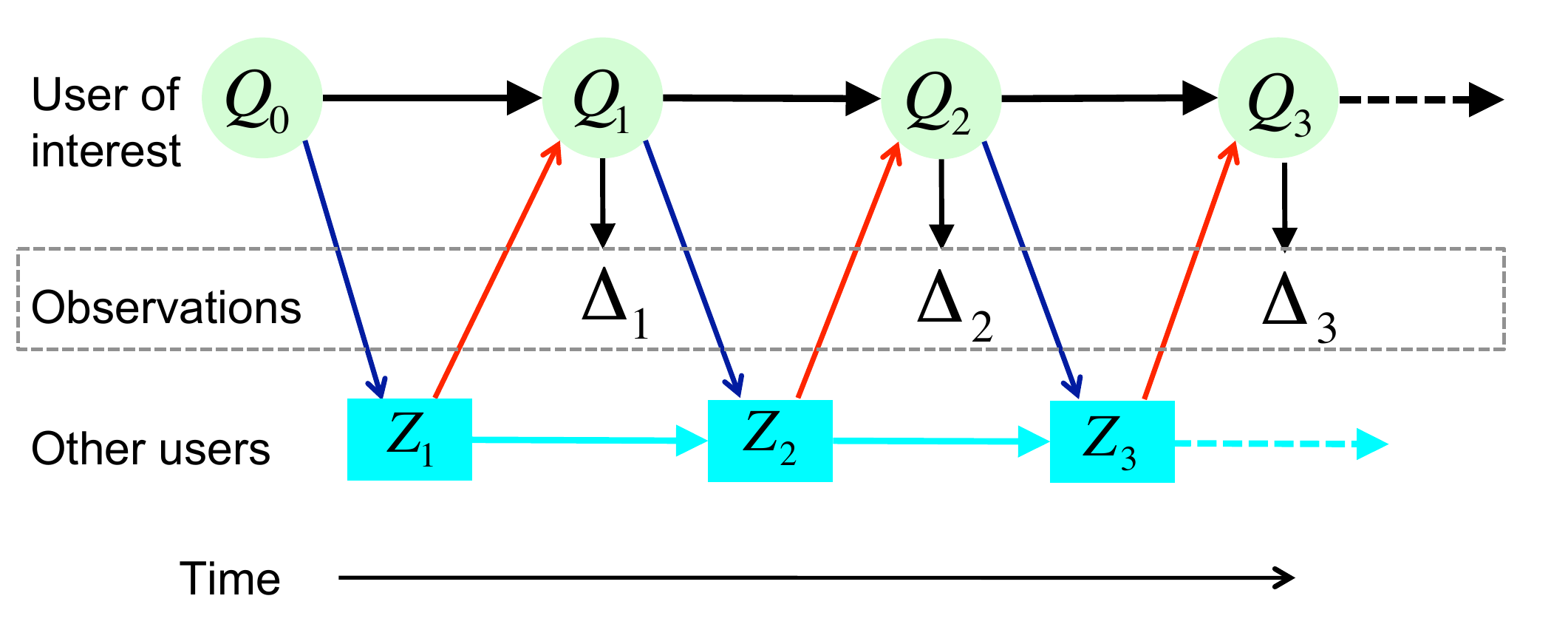}
\end{tabular}
\label{fig1a}
\centering
\end{figure}

\subsection*{Figure 2 - State transition}
Pictorial illustration of state-transition evolution in the proposed model.

\begin{figure}[htb!]
\centering
\begin{tabular}{c}
\includegraphics[height=2.3in,width=3.5in] {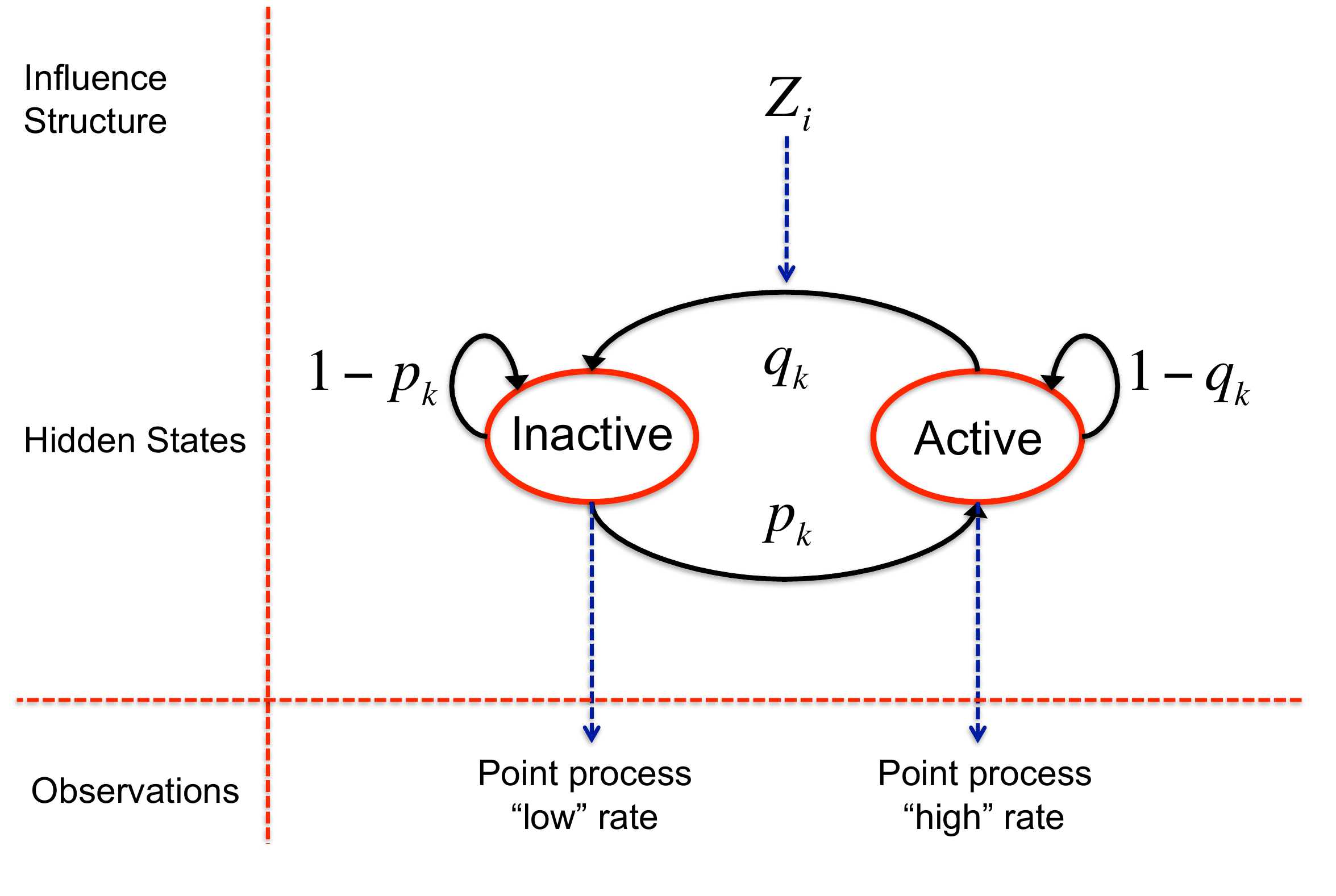}
\end{tabular}
\label{fig1b}
\centering
\end{figure}

\newpage
\subsection*{Figure 3 - Model Validation}
Learned transition probabilities as a function of $\tau$ for User-III
with $n = 1000$.

\begin{figure}[htb!]
\centering
\begin{tabular}{c}
\includegraphics[height=2.5in,width=2.9in] {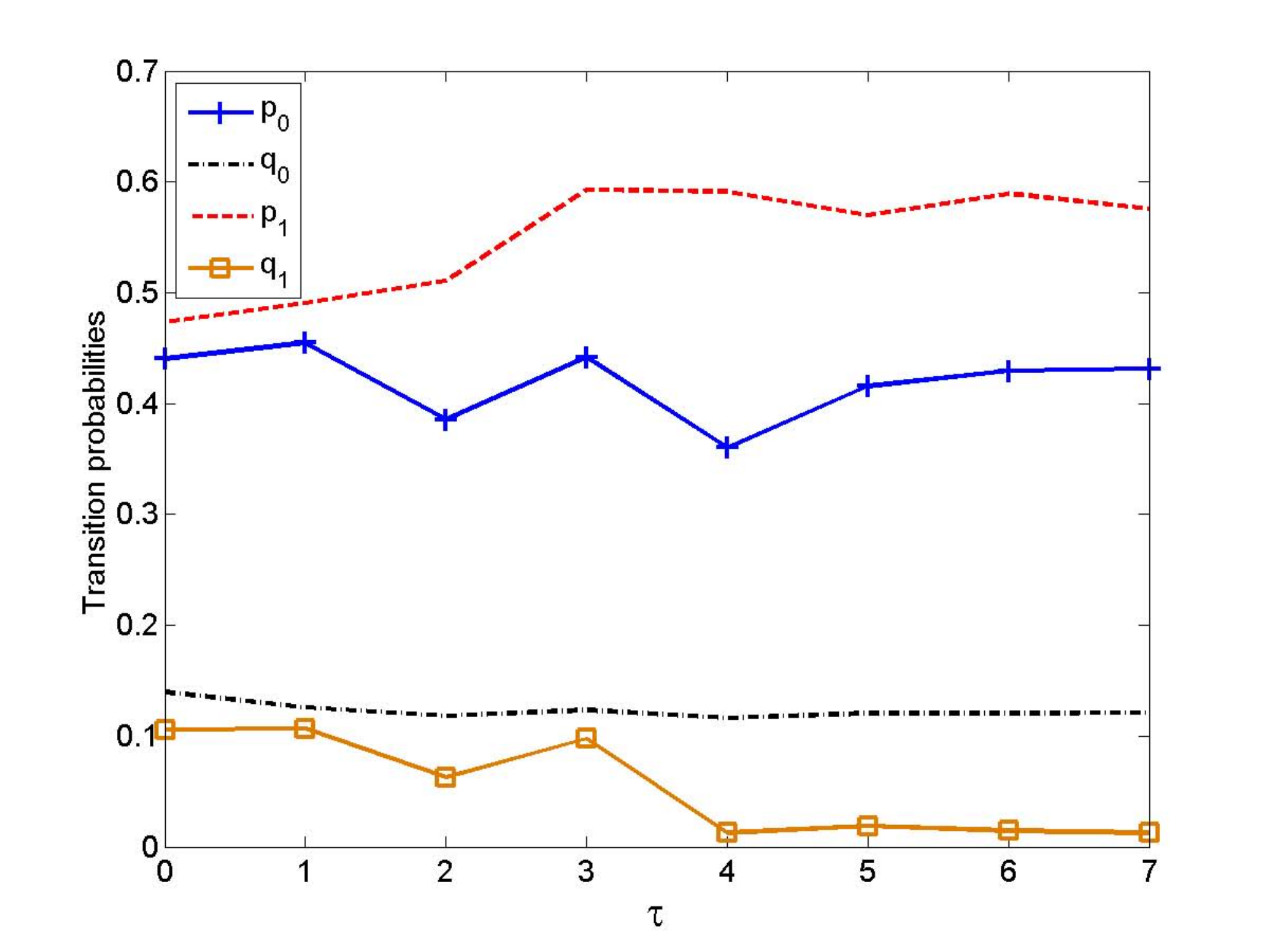}
\end{tabular}
\label{fig_trans}
\centering
\vspace{-5mm}
\end{figure}

\subsection*{Figure 4 - Q-Q Plots}
Q-Q plots of inter-tweet duration in (a)-(d) {\em Active} and (e)-(h) {\em Inactive}
states for User-III under four different models.

\begin{figure*}[htb!]
\begin{center}
\begin{tabular}{cccc}
\includegraphics[height=1.6in,width=1.6in] {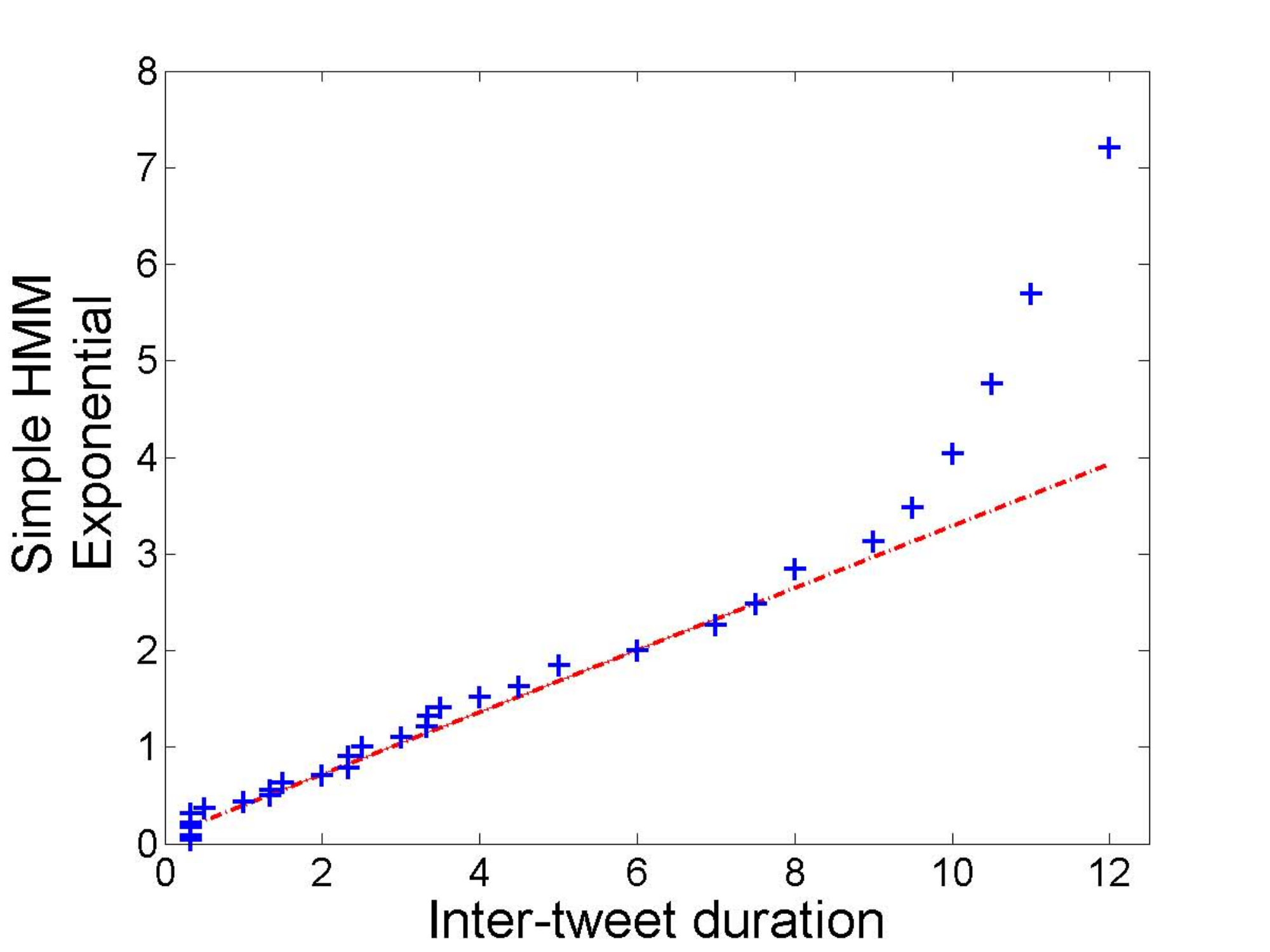}
&
\includegraphics[height=1.6in,width=1.6in] {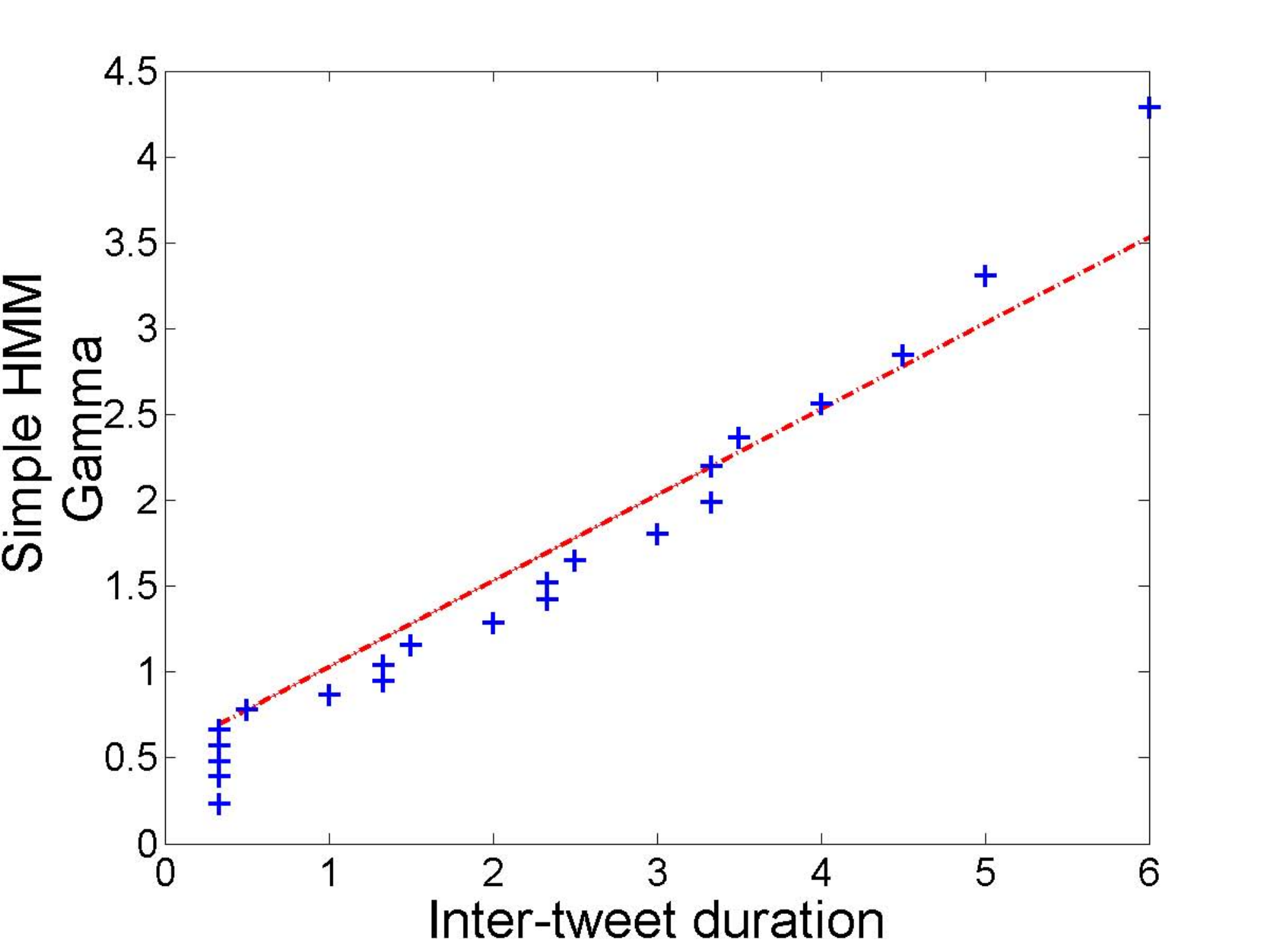}
&
\includegraphics[height=1.6in,width=1.6in] {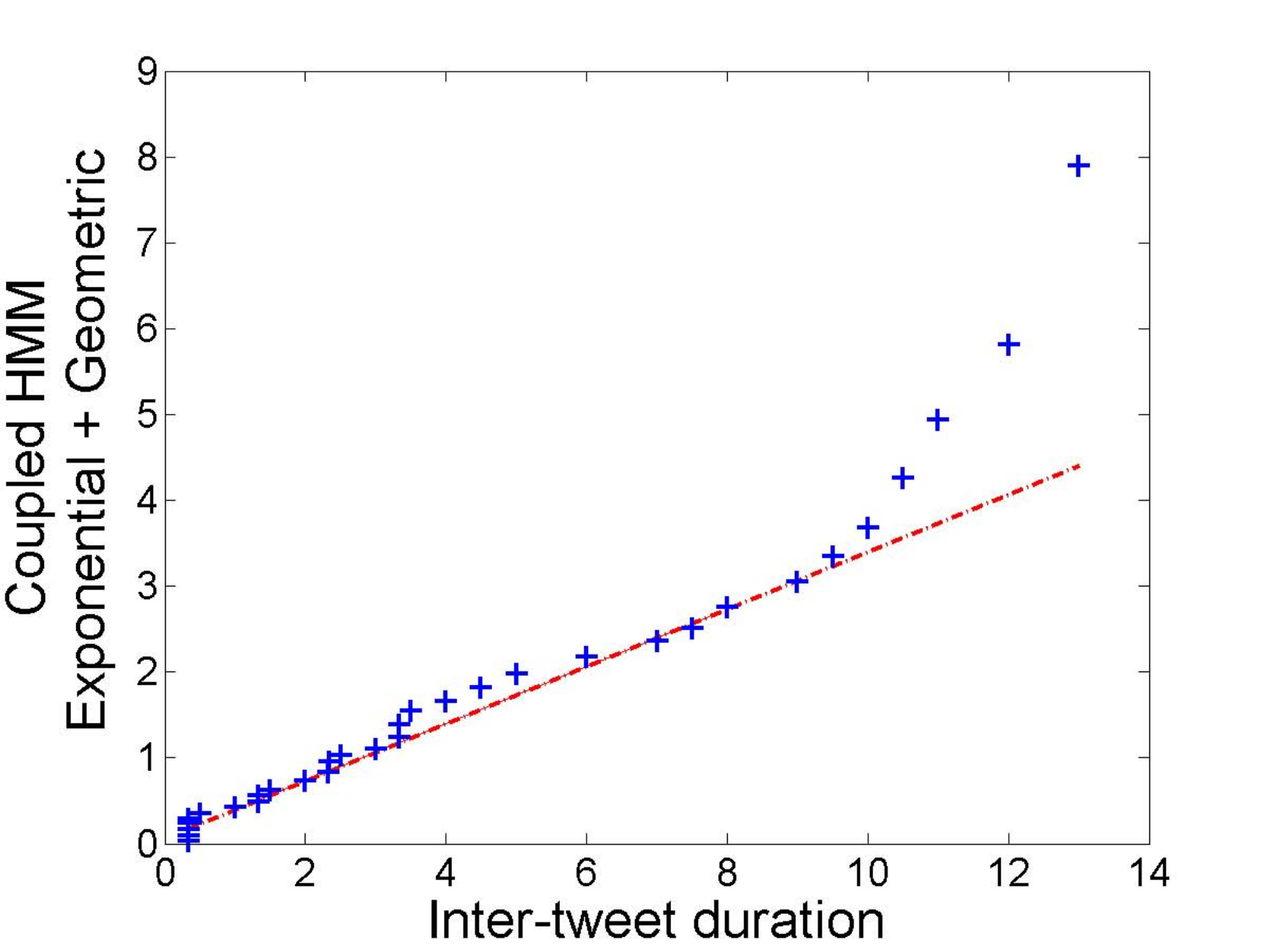}
&
\includegraphics[height=1.6in,width=1.6in] {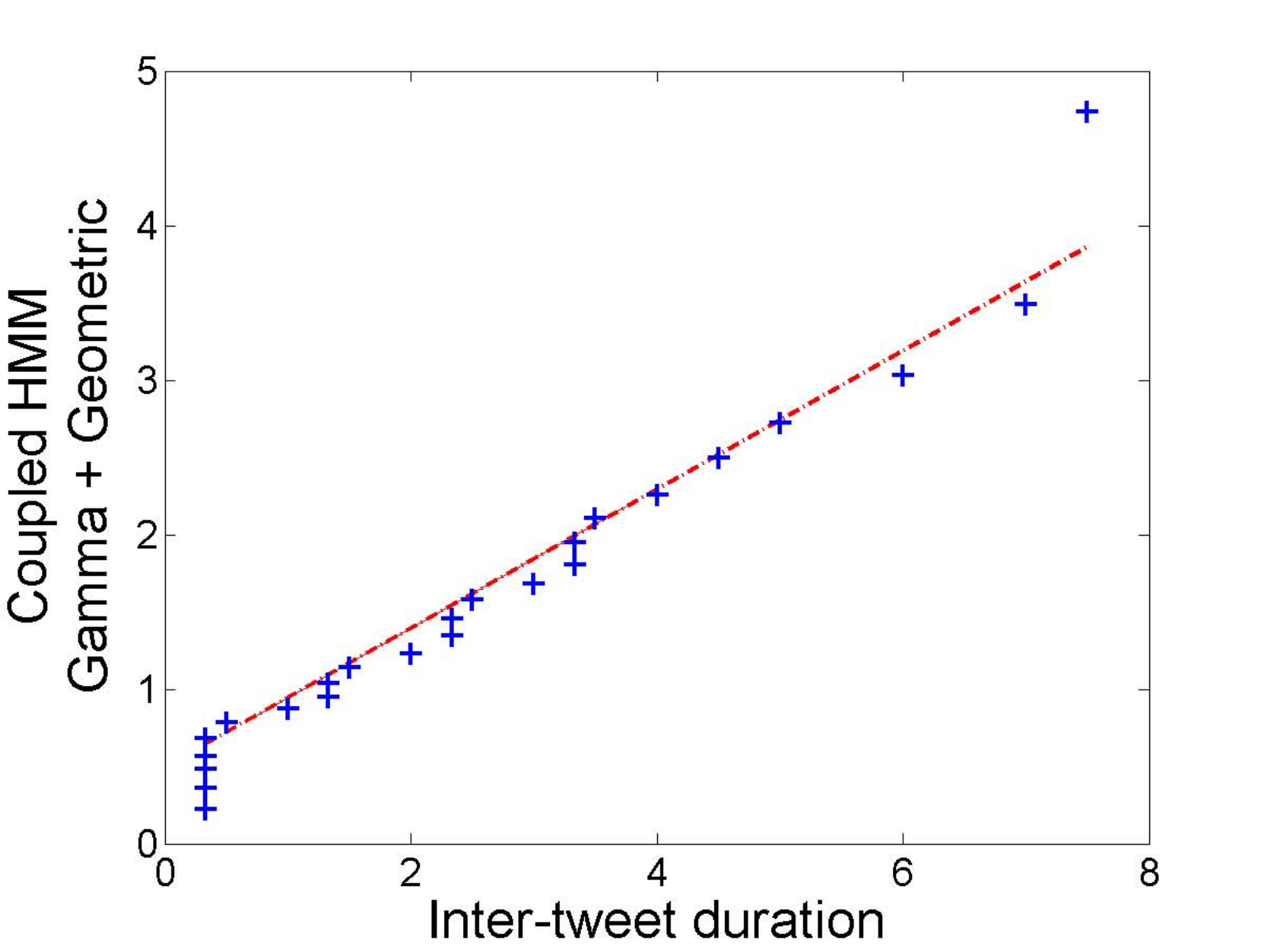}
\\ {\hspace{0.08in}} (a) & {\hspace{0.08in}} (b) & {\hspace{0.08in}} (c) & {\hspace{0.08in}} (d)
\\
\includegraphics[height=1.6in,width=1.6in] {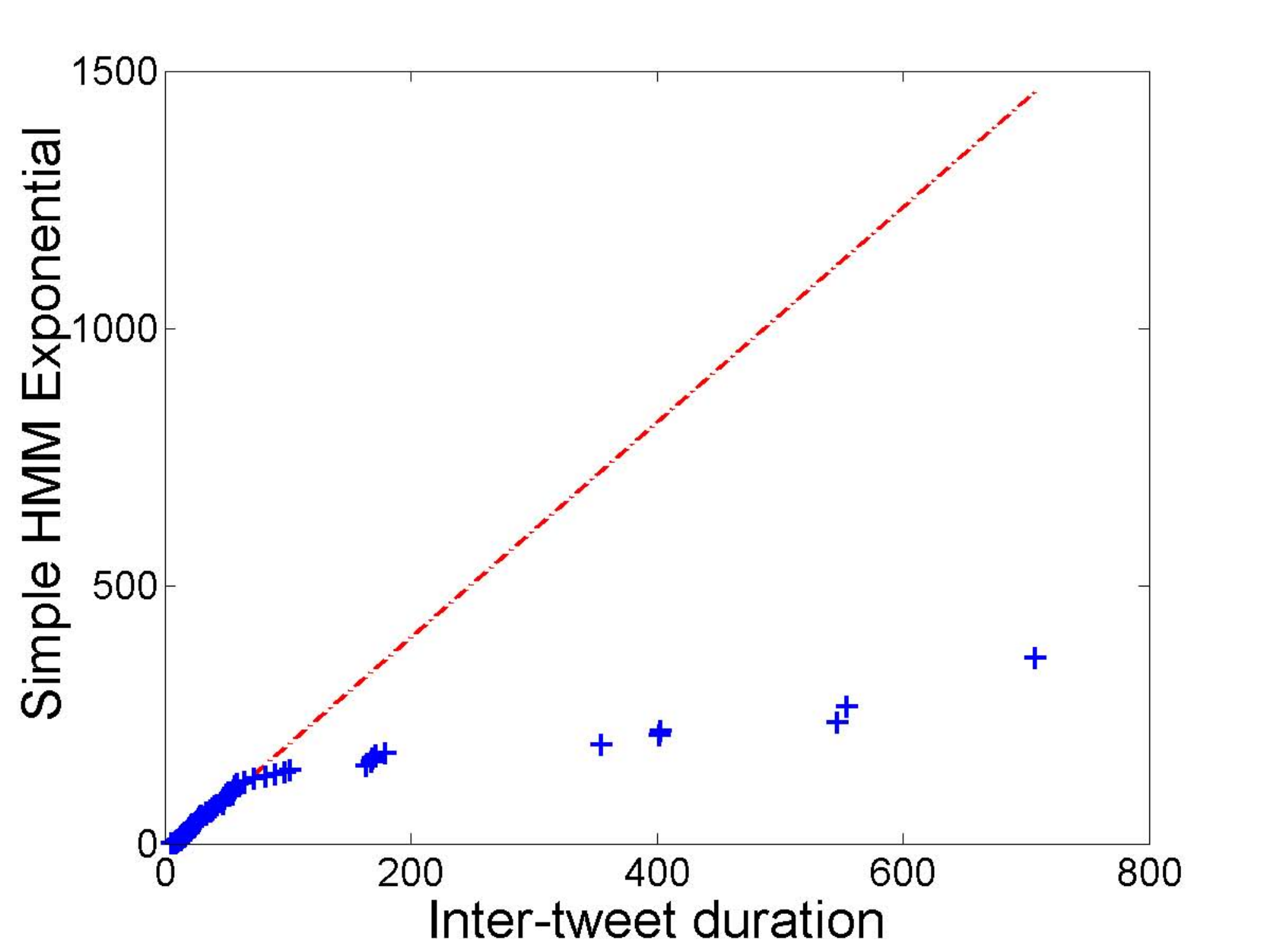}
&
\includegraphics[height=1.6in,width=1.6in] {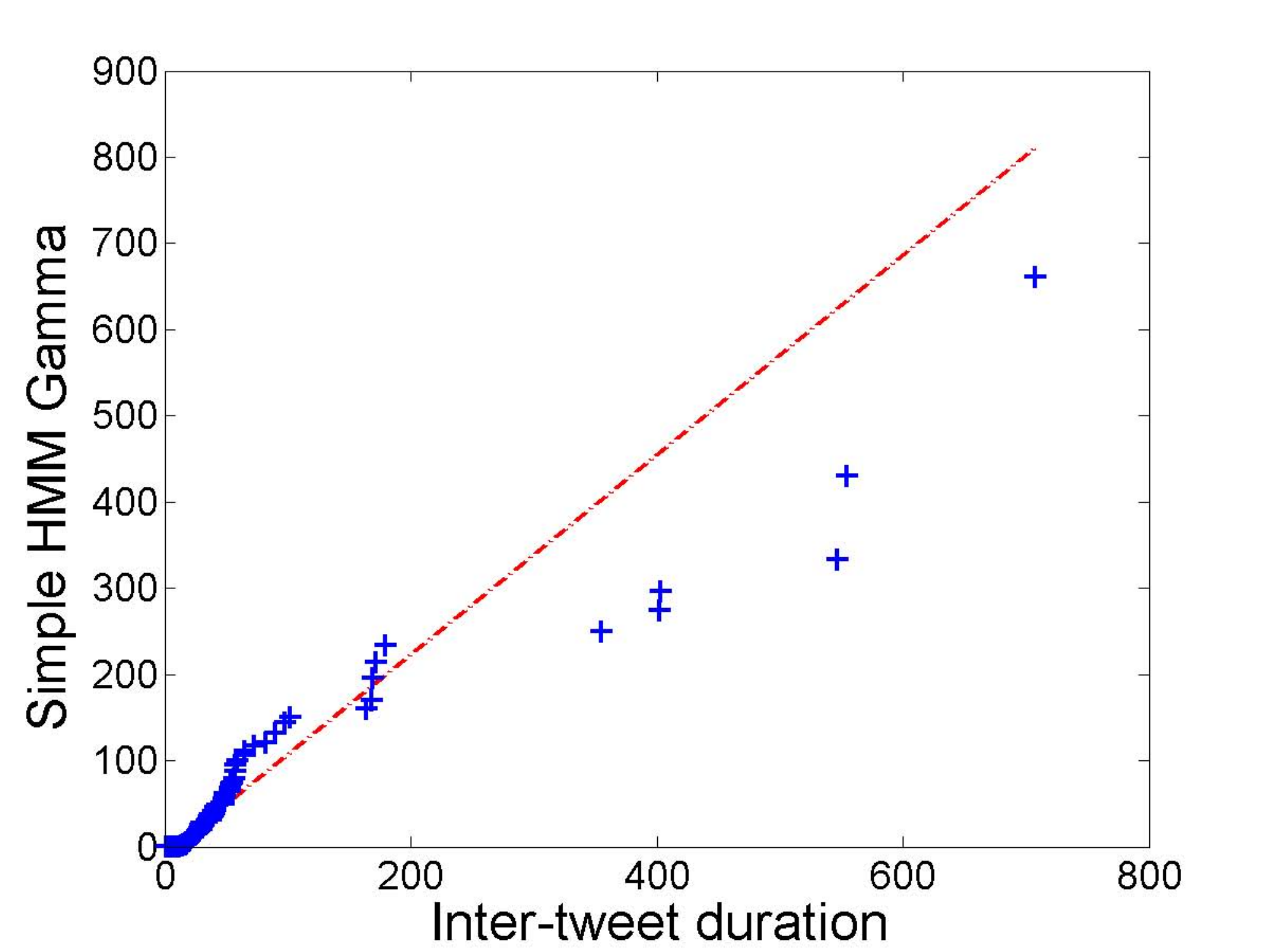}
&
\includegraphics[height=1.6in,width=1.6in] {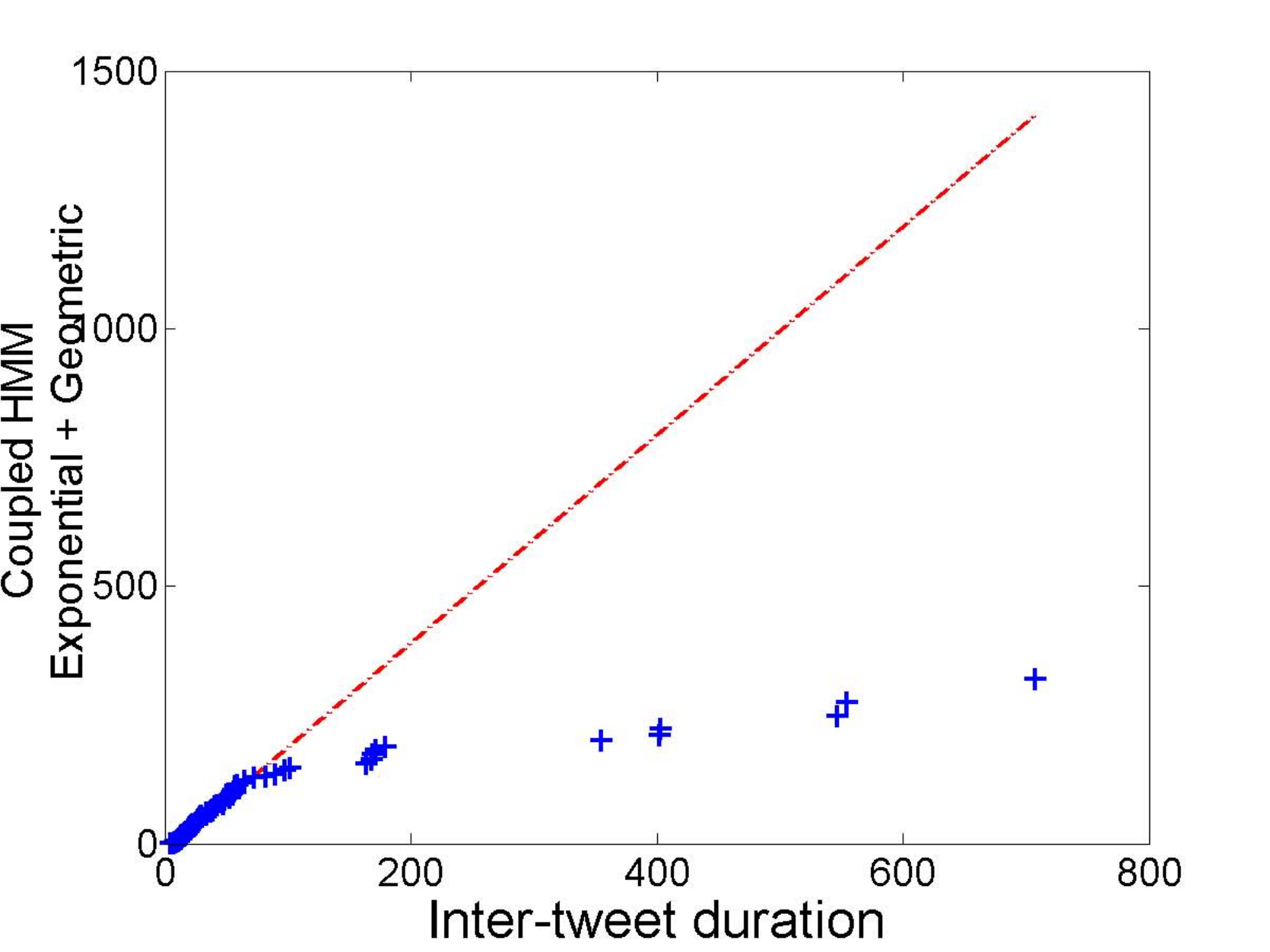}
&
\includegraphics[height=1.6in,width=1.6in] {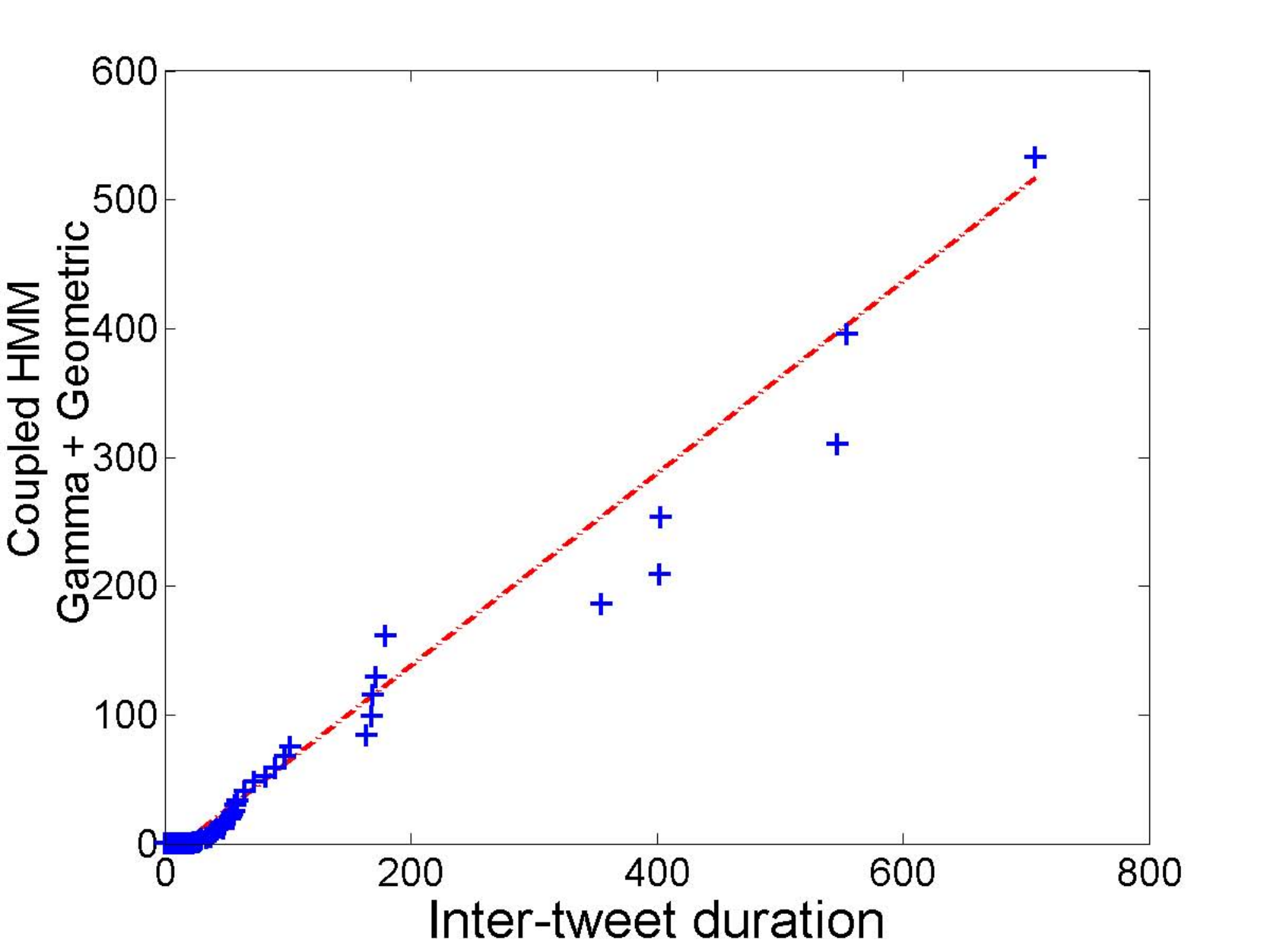}
\\ {\hspace{0.08in}} (e) & {\hspace{0.08in}} (f) & {\hspace{0.08in}} (g) & {\hspace{0.08in}} (h)
\end{tabular}
\label{fig_qqplots}
\end{center}
\vspace{-5mm}
\end{figure*}

\newpage
\subsection*{Figure 5 - Performance Improvement}
AIC improvement for Model {\sf D} relative to Model {\sf B}.

\begin{figure}[htb!]
\centering
\begin{tabular}{c}
\includegraphics[height=2.2in,width=2.7in] {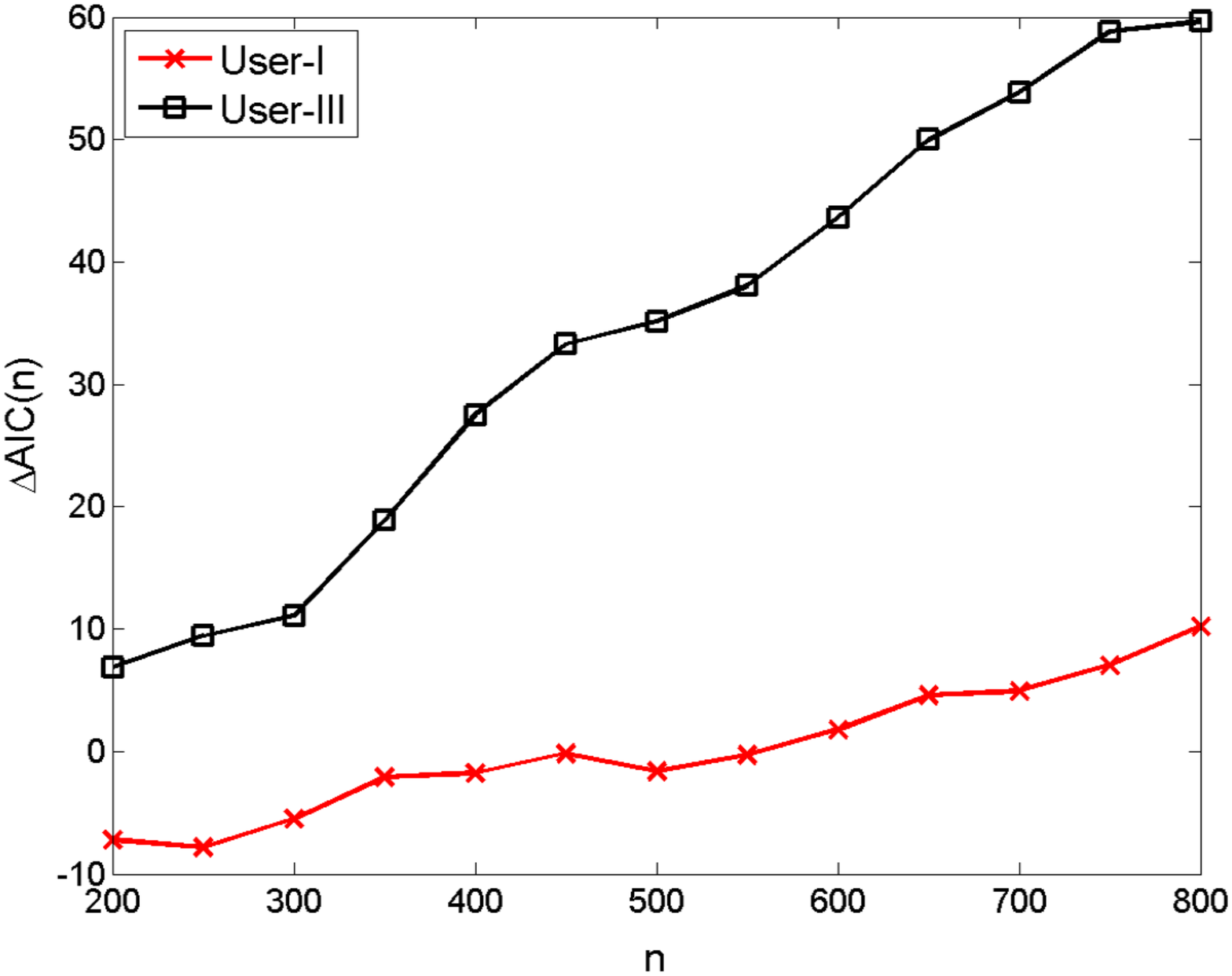}
\end{tabular}
\label{fig_delta}
\centering
\end{figure}

\subsection*{Figure 6 - Histograms}
Histogram of AIC gain with coupled HMM for a corpus of $100$ users with (a)
$n = 500$ and (b) $n = 1000$ observations. (c) Histogram of SMAPE gain with
coupled HMM for $n = 500$.

\begin{figure*}[htb!]
\begin{center}
\begin{tabular}{ccc}
\includegraphics[height=2.1in,width=2.1in] {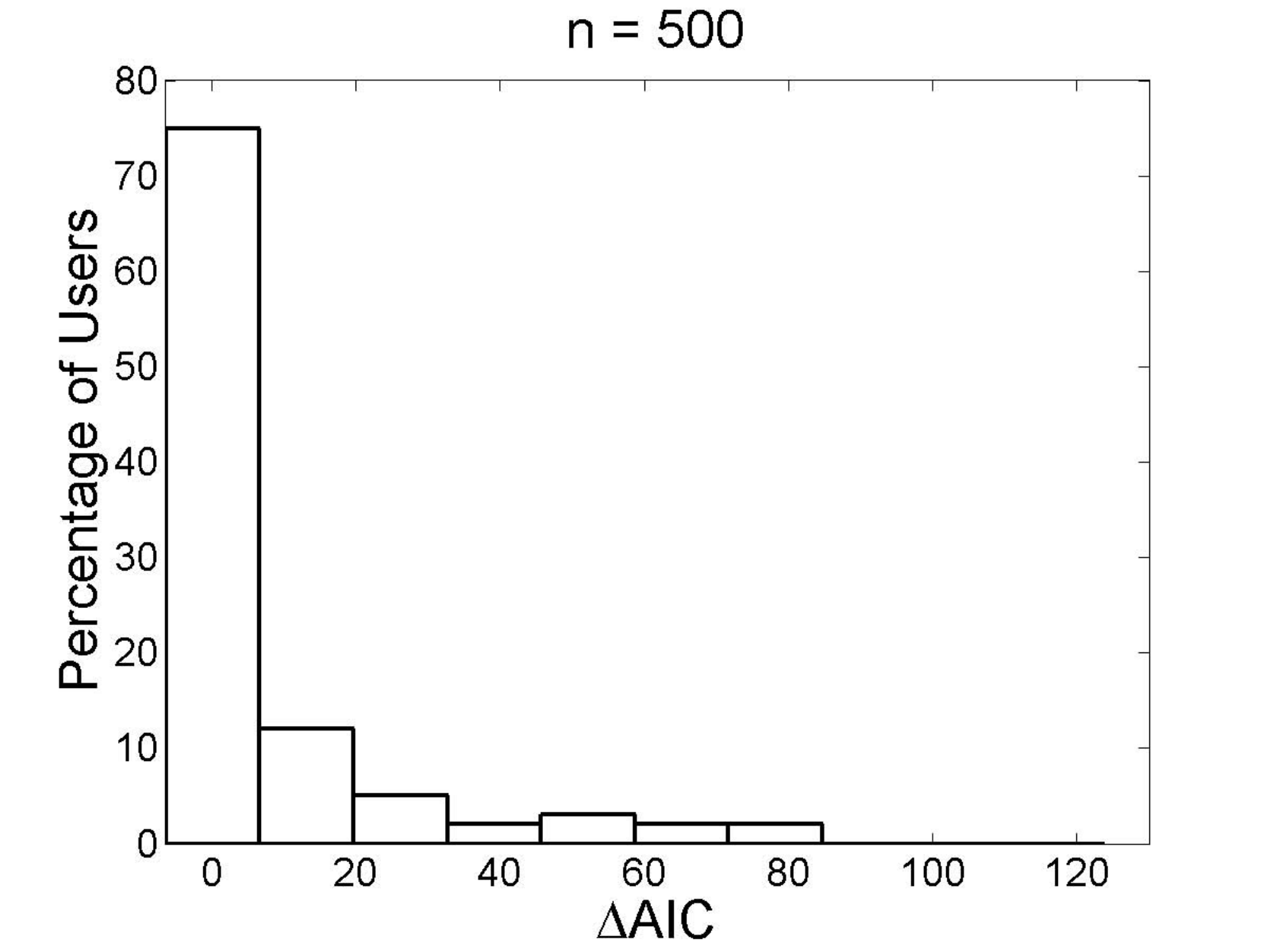}
&
\includegraphics[height=2.1in,width=2.1in] {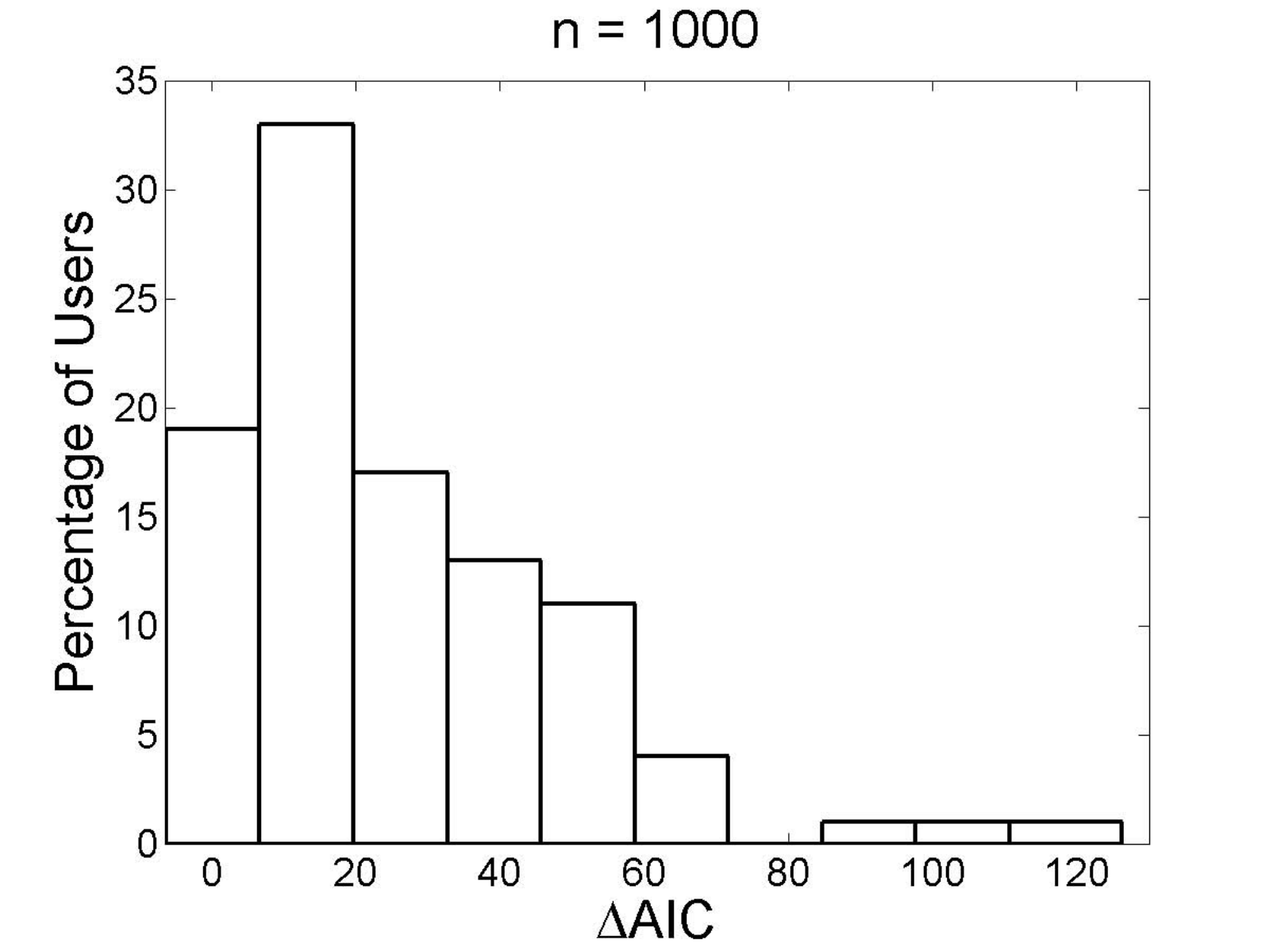}
&
\includegraphics[height=2.1in,width=2.1in] {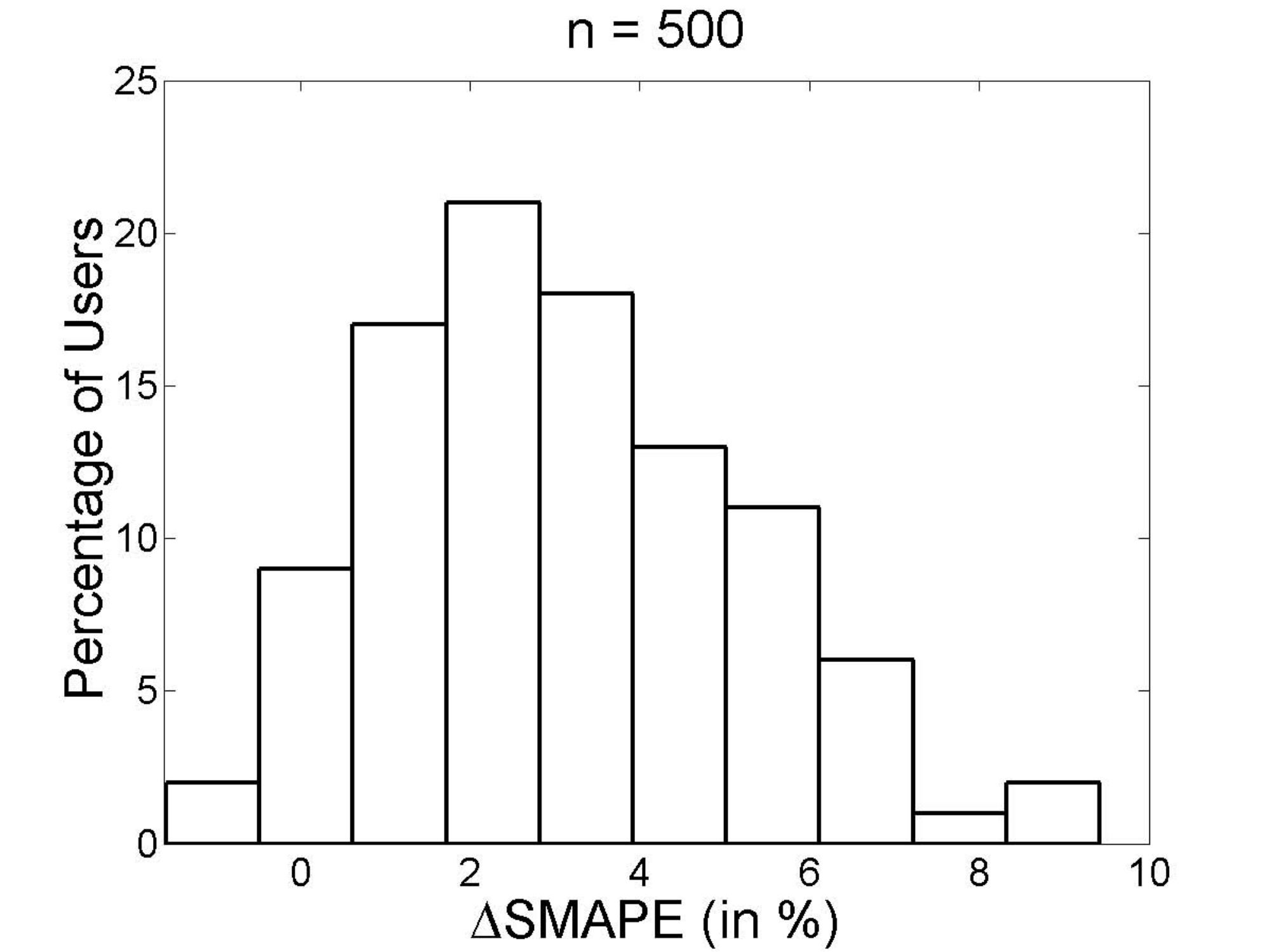}
\\ {\hspace{0.08in}} (a) & {\hspace{0.08in}} (b) & {\hspace{0.08in}} (c)
\end{tabular}
\label{fig_user}
\end{center}
\vspace{-5mm}
\end{figure*}

\newpage
\subsection*{Figure 7 - User Clustering}
Clustering $150$ users into three clusters based on learned model
parameter values, $p_1/p_0$ and $\gamma_1/\gamma_0$.

\begin{figure}[htb!]
\centering
\begin{tabular}{c}
\includegraphics[height=2.5in,width=2.9in] {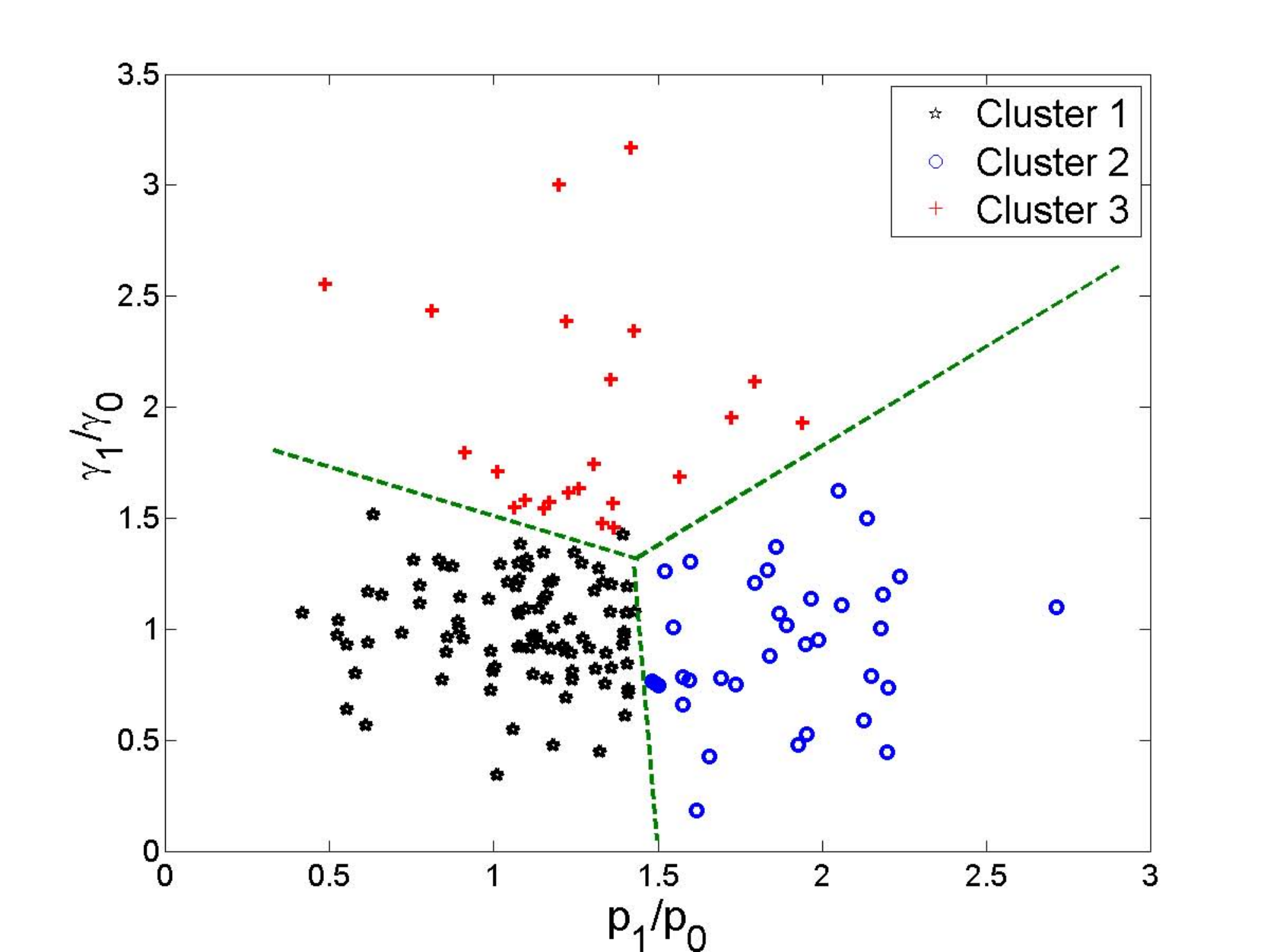}
\end{tabular}
\label{fig2_pre}
\centering
\vspace{-5mm}
\end{figure}
\newpage
\section*{Tables}

\subsection*{Table 1 - Observation Models}
Models for observation given state $Q_i = j, \hsppp j \in \{ 0, 1 \}$.

\begin{table*}[htb!]
\label{table1}
\begin{center}
\begin{tabular}{|c||c|c|c|}
\hline
{\rm Name} & {\rm Density function} ($f_j(\Delta_i)$) &
{\rm Baum-Welch parameter estimate} 
\\ \hline
{\rm Exponential} &
$\rho_j \cdot \exp(  -\rho_j \cdot \Delta_i)$ &
$\widehat{\rho}_j = \frac{ \sum \limits_{i=1}^{N-1}
\big( \xi_{i}(j,0) + 
\xi_{i}(j,1) \big) }
{  \sum \limits_{i=1}^ {N-1} \Delta_i \cdot
\big( \xi_{i}(j,0) + 
\xi_{i}(j,1) \big) }$ \\
\hline
{\rm Gamma} &
$\frac{1}  {\Gamma(k_j) \lambda_j^{k_j} } \cdot
\Delta_i^{k_j - 1} \exp \Big( - \frac{\Delta_i}
{ \lambda_j } \Big)$ &
$\widehat{ k}_j$ and $\widehat{\lambda}_j$ solve:  \hspp
$k_j \lambda_j = \frac{ \sum \limits_{i=1}^{N-1}
\Delta_i \cdot \big( \xi_{i}(j,0) + \xi_{i}(j,1)
\big) } {  \sum \limits_{i=1}^ {N-1}
\big( \xi_{i}(j,0) +  \xi_{i}(j,1) \big) }$
\\
& where $\Gamma(x)$ is the Gamma function &
$\log(\lambda_j) + \Psi(k_j) = \frac{
\sum \limits_{i=1}^{N-1} \log(\Delta_i) \cdot
\big( \xi_{i}(j,0) + \xi_{i}(j,1) \big)  }
{  \sum \limits_{i=1}^ {N-1} \big(
\xi_{i}(j,0) + \xi_{i}(j,1) \big) }$
\\
& & where $\Psi(x) = \frac{\Gamma'(x)}{\Gamma(x)}$ is
the di-gamma function \\
\hline
\ignore{
{\rm Zero-modified} &
$\frac{\gamma_j}{\overline{\Delta}_j}$
if $\Delta_i \leq \overline{\Delta}_j$ &
$\widehat{\gamma}_j =  \frac{ \sum \limits_{i=1}^{N-1}
\indic(\Delta_i \leq \overline{\Delta}_j)
\cdot \big( \xi_{i}(j,0) +\xi_{i}(j,1) \big)  }
{  \sum \limits_{i=1}^ {N-1} \big(
\xi_{i}(j,0) + \xi_{i}(j,1) \big) }$   \\
{\rm exponential} &
$(1 - \gamma_j) \rho_j \exp \Big( - \rho_j
(\Delta_i - \overline{\Delta}_j ) \Big)$ if
$\Delta_i > \overline{\Delta}_j$ &
$\widehat{\rho}_j =
\frac{ \sum \limits_{i=1}^{N-1}
\indic(\Delta_i > \overline{\Delta}_j)
\cdot \big(  \xi_{i}(j,0) +  \xi_{i}(j,1) \big) }
{ \sum \limits_{i=1}^ {N-1}
(\Delta_i - \overline{\Delta}_j) \cdot
\indic(\Delta_i > \overline{\Delta}_j) \cdot
\big( \xi_{i}(j,0) + \xi_{i}(j,1) \big) }$  \\
\hline
& \multicolumn{2}{l|}
{ 
Note: $\indic(\cdot)$ denotes the indicator function of the set under
consideration. }
\\
}
\hline
\end{tabular}
\end{center}
\end{table*}

\subsection*{Table 2 - Influence Structure Models}
Models for influence structure given state $Q_{i-1} = j, \hsppp j
\in \{ 0, 1 \}$.

\begin{table*}[htb!]
\label{table_g}
\begin{center}
\begin{tabular}{|c||c|c|c|}
\hline
{\rm Name} & {\rm Density function} ($g_j(Z_i)$) &
{\rm Baum-Welch parameter estimate}
\\ \hline
{\rm Binary mentions} &
$ {\sf P}(Z_i = k | Q_{i-1} = j) =
\left\{ \begin{array}{cc}
1 - \gamma_j & {\rm if} \hsppp k = 0 \\
\gamma_j & {\rm if} \hsppp k = 1
\end{array} \right.$
& $\widetilde{\gamma}_j =
\frac{ \sum _{ i = 1, \hspp i \hsppp : \hsppp Z_i = 1}^N
\left( \widetilde{\xi}_{i-1}(j,0) + \widetilde{\xi}_{i-1}(j,1) \right)
} { \sum _{ i = 1}^N
\left( \widetilde{\xi}_{i-1}(j,0) + \widetilde{\xi}_{i-1}(j,1) \right) }$
\\ \hline
{\rm No.\ of mentions} & & \\
\cline{1-1}
{\rm Geometric}
& ${\sf P}(Z_i = k | Q_{i-1} = j) =
(1 - \gamma_j) \cdot \gamma_j^k, \hsppp k \geq 0$
& $\widetilde{\gamma}_j = \frac{
\sum_{i = 1}^{N} Z_i \cdot
\left( \widetilde{\xi}_{i-1}(j,0) + \widetilde{\xi}_{i-1}(j,1) \right) }
{ \sum_{i = 1}^{N} (Z_i + 1) \cdot
\left( \widetilde{\xi}_{i-1}(j,0) + \widetilde{\xi}_{i-1}(j,1) \right) }$
\\
\hline
{\rm Poisson}
& ${\sf P}(Z_i = k | Q_{i-1} = j) =
\frac{ \gamma_j^k \cdot \exp(- \gamma_j) } { \Gamma(k + 1)},
\hsppp k \geq 0$
& $\widetilde{\gamma}_j = \frac{
\sum_{i = 1}^{N} Z_i \cdot
\left( \widetilde{\xi}_{i-1}(j,0) + \widetilde{\xi}_{i-1}(j,1) \right) }
{ \sum_{i = 1}^{N}
\left( \widetilde{\xi}_{i-1}(j,0) + \widetilde{\xi}_{i-1}(j,1) \right) }$
\\ \hline
{\rm Shifted zeta}
& ${\sf P}(Z_i = k | Q_{i-1} = j) =
\frac{ (1 + k)^{- \gamma_j}  } { \zeta(\gamma_j) },
\hsppp k \geq 0$
& $\widetilde{\gamma}_j$ solves:
\\
& where $\zeta(x)$ is the Riemann-zeta function &
$\frac{ -\zeta'(\gamma_j)}{\zeta(\gamma_j)} =
\frac{ \sum_{i = 1}^{N} \log(1 + Z_i) \cdot
\left( \widetilde{\xi}_{i-1}(j,0) + \widetilde{\xi}_{i-1}(j,1) \right) }
{ \sum_{i = 1}^{N} \left( \widetilde{\xi}_{i-1}(j,0) +
\widetilde{\xi}_{i-1}(j,1) \right) }$ \\ \hline
\hline
\end{tabular}
\end{center}
\end{table*}

\subsection*{Table 3 - Conditional Dependency Assumptions}
Conditional dependencies of the involved variables in different HMM
architectures.

\begin{table*}[htb!]
\label{table_dep}
\begin{center}
\begin{tabular}{|c||c||c||c||c|}
\hline
{\rm First-Order} &
{\rm IO-HMM}~\cite{bengio} &
{\rm Coupled Factorial} &
{\rm Coupled} &
{\rm Proposed}
\\
{\rm AR-HMM}~\cite{poritz,juang_rabiner1,kenny_lennig}
& & {\rm HMM}~\cite{ghahramani}
& {\rm HMM}~\cite{brand_oliver_pentland,zhong_ghosh}
& {\rm Model} \\
\hline
$\left\{ Q_i , \hsppp \Delta_{i-1} \right\} \longrightarrow \Delta_i$ &
$\left\{ Z_i , \hsppp Q_i \right\} \longrightarrow \Delta_i$ &
$\left\{ Z_i , \hsppp Q_i \right\} \longrightarrow \Delta_i$ &
$Q_i \longrightarrow \Delta_i$ &
$Q_i \longrightarrow \Delta_i$
\\
- &
$Z_{i-1} \longrightarrow Z_i$ &
$Z_{i-1} \longrightarrow Z_i$ &
$\left\{ Q_{i-1}, \hsppp Z_{i-1} \right\} \longrightarrow Z_i$ &
$Q_{i-1} \longrightarrow Z_i$
\\
$Q_{i-1} \longrightarrow Q_i$ &
$\left\{ Q_{i-1}, \hsppp Z_i \right\} \longrightarrow Q_i$ &
$\left\{ Q_{i-1}, \hsppp Z_{i-1} \right\} \longrightarrow Q_i$ &
$\left\{ Q_{i-1}, \hsppp Z_{i-1} \right\} \longrightarrow Q_i$ &
$\left\{ Q_{i-1}, \hsppp Z_i \right\} \longrightarrow Q_i$
\\
\hline
\hline
\end{tabular}
\end{center}
\end{table*}

\newpage
\subsection*{Table 4 - Number of Parameters}
Number of parameters describing different models with $k$ states,
$\ell$ network influence structure levels and an $m$ parameter observation
density in each state. Number of parameters for $k = \ell = 2$ and $m = 1$
or $m = 2$ are also provided.

\begin{table*}[htb!]
\label{table_model}
\begin{center}
\begin{tabular}{|c||c|c|c|c|c|c|c|}
\hline
{\rm No.\ of parameters} & {\rm Trans.\ prob.} &
{\rm Obs.} & {\rm Influence} & {\rm Total}
&
\multicolumn{2}{c|}{ $m$ }
\\
\cline{1-1} \cline{6-7}
{Model $(\downarrow)$} & {\rm matrix} &
{\rm density} & {\rm structure} & & $1$ & $2$
\\
\hline
{\rm Conventional HMM} &
$k(k-1)$ & $km$ & - & $k(k + m - 1)$ & $4$ & $6$ \\
\hline
{\rm First-Order AR-HMM} &
$k(k-1)$ & $k(m+1)$ & - & $k(k + m)$ & $6$ & $8$ \\
\hline
{\rm IO-HMM} & $\ell k (k-1)$ & $\ell k m$ &
$\ell (\ell -1 )$ & $\ell( k (k + m - 1) + \ell - 1)$ & $10$ & $14$  \\
\hline
{\rm Coupled Fact.\ HMM} & $\ell k (k-1)$ & $\ell k m$ &
$\ell (\ell -1 )$ & $\ell( k (k + m - 1) + \ell - 1)$ & $10$ & $14$ \\
\hline
{\rm Coupled HMM}~\cite{brand_oliver_pentland} &
$(\ell + k )(k - 1)$ &
$km$ &
$(\ell + k )(\ell - 1)$ &
$(\ell + k )(\ell + k - 2)$
& $10$ & $12$ \\
& & & & $ + km$ & & \\
\hline
{\rm Coupled HMM}~\cite{zhong_ghosh}
& $2(\ell + k)(k-1)$ & $km$ & $2(\ell + k) (\ell - 1)$
& $2(\ell + k)(\ell + k - 2) + km$
& $18$ & $20$ \\
\hline
{\rm Proposed Model}
& $\ell k(k-1)$ & $km$ & $k(\ell - 1) $ & $k(\ell k + m - 1)$
& $8$ & $10$
\\
\hline
\hline
\end{tabular}
\end{center}
\end{table*}

\newpage
\subsection*{Table 5 - Algorithm Derivation}
Steps in generalized Baum-Welch and Viterbi algorithms.
Here, $a, b$ and $c$ are state variable values and $i$ is the
time-index.

\begin{figure*}[!htbp]
\normalsize
{\bf \underline{Steps in generalized Baum-Welch algorithm:}}
\begin{eqnarray}
\label{ref1_eqn_begin}
\widetilde{\xi}_i(a,b) & = &  \frac{\widetilde{ \alpha}_i(a) \hsppp
{\sf P}(Z_{i+1}|Q_i = a) \hsppp {\sf P}(Q_{i+1} = b | Q_i = a, Z_{i+1})
\hsppp {\sf P}(\Delta_{i+1} | Q_{i+1} = b) \hsppp
\widetilde{\beta}_{i+1}(b) } {
\sum \limits_{a,b} \widetilde{\alpha}_i(a) \hsppp {\sf P}(Z_{i+1}|Q_i = a)
\hsppp {\sf P}(Q_{i+1} = b | Q_i = a, Z_{i+1}) \hsppp {\sf P}(\Delta_{i+1}
| Q_{i+1} = b) \hsppp \widetilde{\beta}_{i+1}(b) }, \hspp {\rm where}
\\
\label{eq_alpha1}
\widetilde{\alpha}_i(a) & \triangleq &
{\sf P}(\Delta_1^i, Z_1^i, Q_i = a), \hspp 2 \leq i \leq N-1
\\
\label{eq_alpha2}
& = & \left[ \sum_b \widetilde{\alpha}_{i-1}(b) \hsppp
{\sf P}(Z_i |Q_{i-1} = b) \hsppp {\sf P}(Q_i = a | Q_{i-1} = b, Z_i) \right]
\hsppp {\sf P}(\Delta_i | Q_i = a)
\\
\label{eq_alpha3}
\widetilde{\alpha}_1(a) & = & \left[ \sum_b \pi_b \hsppp
{\sf P}(Z_1 | Q_0 = b) \hsppp {\sf P} (Q_1 = a | Q_0 = b, Z_1) \right] \hsppp
{\sf P}(\Delta_1 |Q_1 = a) 
\\
\widetilde{\beta}_{i}(b) & \triangleq &
{\sf P}( \Delta_{i+1}^N, Z_{i+1}^N | Q_i = b), \hspp
2 \leq i \leq N-1
\\ & = & {\sf P} (Z_{i+1} |Q_i = b) \hsppp \left[
\sum_c {\sf P}(Q_{i+1} = c | Q_i = b, Y_{i+1}) \hsppp
{\sf P}( \Delta_{i+1} | Q_{i+1} = c) \hsppp
\widetilde{\beta}_{i+1} ( c ) \right] 
\\
\widetilde{\beta}_N(b) & = & 1. 
\label{ref1_eqn_end}
\end{eqnarray}
\vspace*{2pt}
\hrulefill

{\bf \underline{Steps in generalized Viterbi algorithm:}}
\begin{eqnarray}
\label{ref2_eqn_begin}
\delta_1(a) & \triangleq &
{\sf P}(Q_1 = a, \Delta_1, Z_1 | \boldsymbol\lambda )
\\ & = & \left[ \sum_b {\sf P}(Q_0 = b) \hsppp {\sf P} (Z_1 | Q_0 = b)
\hsppp {\sf P}(Q_1 = a | Q_0 = b, Z_1) \right] \cdot
{\sf P}(\Delta_1 | Q_1 = a)
\\ \phi_1(a) & = & 0 
\\ \delta_{i+1}(a) & \triangleq & \max \limits_{Q_1^i}
\Big[ {\sf P}(Q_1^i, Q_{i+1} = a,
\Delta_1^{i+1} , Z_1^{i+1} | \boldsymbol\lambda ) \Big], \hspp
1 \leq i \leq N-1 
\\
& = & \max_b \Big[ \delta_i(b) \hsppp {\sf P}(Z_{i+1} | Q_i = b)
\hsppp {\sf P}(Q_{i+1} = a |Q_i = b, Z_{i+1} ) \Big]
\hsppp {\sf P}(\Delta_{i+1} | Q_{i+1} = a) 
\\ \phi_{i+1}(a) & = &
\argmax_b \Big[ \delta_i(b) \hsppp {\sf P}(Z_{i+1} | Q_i = b)
\hsppp {\sf P}(Q_{i+1} = a |Q_i = b, Z_{i+1} ) \Big]. 
\label{ref2_eqn_end}
\end{eqnarray}
\vspace*{2pt}
\hrulefill
\end{figure*}

\newpage
\subsection*{Table 6 - Conditional Correlation Coefficient}
Conditional correlation coefficient
$\rho(Z_i, \hsppp Z_{i-1} | Q_{i-1})$ for Users-I to III in {\em Active}
and {\em Inactive} states.

\begin{table}[htb!]
\label{table_corr}
\begin{center}
\begin{tabular}{|c||c|c|c|c|c|c|}
\hline
& \multicolumn{3}{c|}{$Q_{i-1} = 1$}
& \multicolumn{3}{c|}{$Q_{i-1} = 0$}
\\  \hline
{\rm User ($\downarrow$)} & {\rm No.\ of samples}
& {\rm Corr.\ coef.} & $p$-value &
{\rm No.\ of samples} & {\rm Corr.\ coef.} &
$p$-value
\\ \hline \hline
{\rm User}-I & $660$ & $0.1127$ & $0.0037$ & $138$ & $-0.0244$ & $0.7762$
\\ \hline
{\rm User}-II & $909$ & $0.0590$ & $0.0753$ & $89$ & $-0.0540$ & $0.6155$
\\ \hline
{\rm User}-III & $844$ & $-0.0459$ & $0.1832$ & $154$ & $0.1306$ & $0.1065$
\\ \hline
\hline
\end{tabular}
\end{center}
\end{table}

\subsection*{Table 7 - ${\rm AIC}$ Comparison for User-I}
{\rm AIC} scores for a typical user with different model fits. The
bolded figures correspond to the model with the best fit.

\begin{table}[htb!]
\label{table_aic_bic_raza_shah}
\begin{center}
\begin{tabular}{|c||c|c|c|c|}
\hline
& \multicolumn{3}{c|}{ {\rm AIC}  }
\\ \hline
{\rm Model ($\downarrow$)} & $n = 100$ & $n = 250$ & $n = 500$ & $n = 750$
\\ \hline
\multicolumn{5}{c}
{\rm Observation density: Exponential}
\\  \hline
{\rm Poisson process model} & $1017.71$ & $2711.88$ & $5157.41$ & $7556.58$
\\ \hline
{\rm Conventional two-state HMM}
& $835.42$ & $2155.51$ & $3852.15$ & $5630.03$
\\ \hline
{\rm Coupled HMM, Binary mention}
& $844.19$ & $2169.88$ & $3867.18$ & $5658.23$
\\ \hline
{\rm Coupled HMM, No. mentions (Geometric)}
& $840.65$ & $2163.20$ & $3853.10$ & $5621.40$
\\ \hline
{\rm Coupled HMM, No. mentions (Poisson)}
& $840.96$ & $2163.54$ & $3853.57$ & $5621.90$
\\ \hline
{\rm Coupled HMM, No. mentions (Shifted zeta)}
& $841.93$ & $2163.39$ & $3853.83$ & $5623.03$
\\ \hline
{\rm Coupled HMM, Social network traffic (Geometric)}
& $836.22$ & $2156.11$ & $3951.16$ & $5789.35$
\\ \hline
\multicolumn{5}{c}
{\rm Observation density: Gamma}
\\ \hline
{\rm Renewal process model} & $879.02$ & $2284.01$ & $4214.99$ & $6230.80$
\\ \hline
{\rm Conventional two-state HMM}
& ${\bf 830.66}$ & ${\bf 2134.12}$ & $3806.88$ & $5578.28$
\\ \hline
{\rm Coupled HMM, Binary mention}
& $838.81$ & $2148.02$ & $3816.54$ & $5595.53$
\\ \hline
{\rm Coupled HMM, No. mentions (Geometric)}
& $836.36$ & $2141.94$ & $3808.53$ & ${\bf 5571.18}$
\\ \hline
{\rm Coupled HMM, No. mentions (Poisson)}
& $837.42$ & $2142.66$ & $3810.11$ & $5572.50$
\\ \hline
{\rm Coupled HMM, No. mentions (Shifted zeta)}
& $837.76$ & $2142.13$ & $3809.29$ & $5584.69$
\\ \hline
{\rm Coupled HMM, Social network traffic (Geometric)}
& $834.81$ & $2149.96$ & ${\bf 3797.65}$ & $5583.40$
\\ \hline \hline
\end{tabular}
\end{center}
\end{table}

\newpage
\subsection*{Table 8 - ${\rm AIC}$ Comparison for User-II}
{\rm AIC} scores for a typical user with different model fits. The
bolded figures correspond to the model with the best fit.

\begin{table}[htb!]
\label{table_aic_bic_mansoorzia}
\begin{center}
\begin{tabular}{|c||c|c|c|}
\hline
& \multicolumn{3}{c|}{ {\rm AIC}  }
\\ \hline
{\rm Model ($\downarrow$)} & $n = 100$ & $n = 500$ & $n = 1000$
\\ \hline
\multicolumn{4}{c}
{\rm Observation density: Exponential}
\\  \hline
{\rm Poisson process model} & $698.48$ & $3444.09$ & $6837.57$
\\ \hline
{\rm Conventional two-state HMM} & $504.20$ & $2373.26$ & $4343.94$
\\ \hline
{\rm Coupled HMM, Binary mention}
& $517.35$ & $2383.19$ & $4354.92$
\\ \hline
{\rm Coupled HMM, No. mentions (Geometric)}
& $504.62$ & $2364.04$ & $4333.98$
\\ \hline
{\rm Coupled HMM, No. mentions (Poisson)}
& $504.42$ & $2363.96$ & $4333.78$
\\ \hline
{\rm Coupled HMM, No. mentions (Shifted zeta)}
& $504.76$ & $2363.94$ & $4333.80$
\\ \hline
{\rm Coupled HMM, Social network traffic (Geometric)}
& $489.87$ & $2391.80$ & $4406.64$ \\ \hline
\multicolumn{4}{c}
{\rm Observation density: Gamma}
\\ \hline
{\rm Renewal process model} & $613.98$ & $2960.90$ & $5691.82$
\\ \hline
{\rm Conventional two-state HMM} & $500.68$ & $2334.20$ & $4261.41$
\\ \hline
{\rm Coupled HMM, Binary mention}
& $511.86$ & $2339.44$ & $4265.43$
\\ \hline
{\rm Coupled HMM, No. mentions (Geometric)}
& $501.72$ & $2320.02$ & $4257.15$
\\ \hline
{\rm Coupled HMM, No. mentions (Poisson)}
& $501.39$ & ${\bf 2319.73}$ & ${\bf 4256.61}$
\\ \hline
{\rm Coupled HMM, No. mentions (Shifted zeta)}
& $501.92$ & $2320.04$ & $4256.93$
\\ \hline
{\rm Coupled HMM, Social network traffic (Geometric)}
& ${\bf 487.70}$ & $2336.06$ & $4283.44$ \\ \hline
\hline
\end{tabular}
\end{center}
\end{table}

\newpage
\subsection*{Table 9 - ${\rm AIC}$ Comparison for User-III}
{\rm AIC} scores for an extreme case of a highly active user with
different model fits. The bolded figures correspond to the model with
the best fit.

\begin{table}[htb!]
\label{table_aic_bic_marvi}
\begin{center}
\begin{tabular}{|c||c|c|c|}
\hline
& \multicolumn{3}{c|}{ {\rm AIC}  }
\\ \hline
{\rm Model ($\downarrow$)} & $n = 100$ & $n = 500$ & $n = 1000$
\\ \hline
\multicolumn{4}{c}
{\rm Observation density: Exponential}
\\  \hline
{\rm Poisson process model} & $653.73$ & $3141.36$ & $6250.35$
\\ \hline
{\rm Conventional two-state HMM} & $479.31$ & $2359.48$ & $4611.80$
\\ \hline
{\rm Coupled HMM, Binary mention}
& $495.63$ & $2391.71$ & $4677.30$ 
\\ \hline
{\rm Coupled HMM, No. mentions (Geometric)}
& $478.69$ & $2337.30$ & $4552.67$ 
\\ \hline
{\rm Coupled HMM, No. mentions (Poisson)}
& $483.11$ & $2339.65$ & $4682.30$ 
\\ \hline
{\rm Coupled HMM, No. mentions (Shifted zeta)}
& $490.09$ & $2370.70$ & $4753.87$ 
\\ \hline
{\rm Coupled HMM, Social network traffic (Geometric)}
& $499.81$ & $2402.47$ & $4663.76$ \\ \hline
\multicolumn{4}{c}
{\rm Observation density: Gamma}
\\ \hline
{\rm Renewal process model} & $585.78$ & $2830.55$ & $5594.71$
\\ \hline
{\rm Conventional two-state HMM} & $476.39$ & $2290.53$ & $4440.57$
\\ \hline
{\rm Coupled HMM, Binary mention}
& $485.79$ & $2318.99$ & $4494.15$ 
\\ \hline
{\rm Coupled HMM, No. mentions (Geometric)}
& ${\bf 473.56}$ & ${\bf 2262.10}$ & ${\bf 4394.30}$
\\ \hline
{\rm Coupled HMM, No. mentions (Poisson)}
& $489.48$ & $2262.22$ & $4521.20$ 
\\ \hline
{\rm Coupled HMM, No. mentions (Shifted zeta)}
& $484.44$ & $2300.41$ & $4469.11$ 
\\ \hline
{\rm Coupled HMM, Social network traffic (Geometric)}
& $485.73$ & $2313.83$ & $4458.81$ \\ \hline
\hline
\end{tabular}
\end{center}
\end{table}

\subsection*{Table 10 - ${\rm AIC}$ Improvement}
$\Delta {\tt AIC}$ for Users-I to III with different $n$ values.

\begin{table}[htb!]
\label{table_delta_aic}
\begin{center}
\begin{tabular}{|c||cc|cc|cc|}
\hline
{\rm User ($\downarrow$)} &
\multicolumn{2}{c|}{Small $n$} &
\multicolumn{2}{c|}{Moderate $n$} &
\multicolumn{2}{c|}{Large $n$}
\\  \hline
{\rm User}-I & $-5.70$ & $(n = 100)$ & $-1.65$ & $(n = 500)$ &
$7.10$ & $(n = 750)$ \\ \hline
{\rm User}-II & $-1.04$ & $(n = 100)$ & $14.18$ & $(n = 500)$ &
$4.26$ & $(n = 1000)$ \\ \hline
{\rm User}-III & $2.83$ & $(n = 100)$ & $28.43$ & $(n = 500)$ &
$46.27$ & $(n = 1000)$ \\ \hline
\hline
\end{tabular}
\end{center}
\end{table}

\newpage
\subsection*{Table 11 - Sample Activity Listing}
Sample activity of a typical user in Cluster 2.

\begin{table}[htb!]
\label{typical_act_cluster2}
\begin{center}
\begin{tabular}{|l|l|l|}
\hline
Posted by & Intended for & Text content \\ \hline \hline
@Friend & @Cluster-2-user & yeh.watching \\ \hline
@Cluster-2-user & & RT @Non-Friend: In the U.S., you can text ``FLOOD'' to 27722
\\
& & to donate \$10 to the \#Pakistan Relief Fund. [URL] \#helppakistan
\\ \hline
@Cluster-2-user & @Friend & :D
\\ \hline
@Cluster-2-user & & Is it mandatory for Saeed Ajmal to bowl on short ball in
every \\
& &  over \#fail \#PakCricket
\\ \hline
@Friend & @Cluster-2-user & i think afridi should go for gull
\\ \hline
@Cluster-2-user & & Believe me when I say that I predicted that shoaib
will get yardy \\ & & in this over :D \#PakCricket
\\ \hline
@Cluster-2-user & & That was a khooooni yorker by Umer \#Gull Waqar ki yaad
\\ & & aa gai \#PakCricket
\\ \hline
@Friend & & RT @Cluster-2-user: 300 not out by Boom Boom [URL]
@Friend1 \\ & &  @Friend2 ... @Friend8
\\ \hline
@Cluster-2-user & & Baba dam darood aka Muhammad Yousaf has taken an
\\ & & un believable catch \#PakCricket \\ \hline
@Friend & @Cluster-2-user & ;-) \\ \hline
@Friend & & RT @Cluster-2-user: \#blog post 300 not out by Boom Boom
\\ & & [URL] \#PakCricket \#BoomBoom \#Afridi
\\ \hline
@Friend & @Cluster-2-user & now what? another scandal against
\#pakcricket \#Pakistan \\ & &  for winning the match
\\ \hline
@Cluster-2-user & @Friend & i think the scandal is going to be damaging bat
of Swan on the \\ & &  yorker attempt b Shabby \#PakCricket
\\ \hline
@Cluster-2-user & & RT @Friend: There's always second chances... it
just depends on \\  & &  how hard you fight for it !!
\\ \hline
@Cluster-2-user & & RT @Friend: dedicated to men in Green hope this streak
continues \\ & & and we we the coming one as well.... [URL]
\\ \hline
@Cluster-2-user & & Hay Jazba Janoon to Himmat na haar... the formula
behind \\ & & the success of cricket team \#PakCricket \\ \hline \hline 		
\end{tabular}
\end{center}
\end{table}

\newpage
\subsection*{Table 12 - Sample Argument}
Sample argument between two typical users in Cluster 2.

\begin{table}[htb!]
\label{argument_cluster2}
\begin{center}
\begin{tabular}{|l|l|l|}
\hline
Posted by & Intended for & Text content \\ \hline \hline
@Cluster-2-user-2 & @Cluster-2-user-1 & quick question...is @salmantaseer
one of the bad guys?? i thought \\
& & the PPP was the lesser of the evils over there..
\#justwondering \\ \hline 	
@Cluster-2-user-1 & @Cluster-2-user-2 & PPP lesser of the evils?! Hahahaha
everyone working under Zardari \\
& & is as evil as Bieber can ever dream of being. \\ \hline
@Cluster-2-user-2 & @Cluster-2-user-1 & aww..but sherry rehman is so sweet..and
i love the rehman malik \\
& & hairdo..and SMQ is so totally suave..btw who are the good guys?? \\ \hline 	
@Cluster-2-user-1 & @Cluster-2-user-2 & wow man! You're kidding, right?
\\ \hline
@Cluster-2-user-2 & @Cluster-2-user-1 & man!!why do i get the feeling that
i just said like a totally not cool \\ && thing..	 \\ \hline
@Cluster-2-user-1 & @Cluster-2-user-2 &
Err yes. Totally uncool :P \\ \hline
@Cluster-2-user-2 & @Cluster-2-user-1 & so no good guys huh??and btw yes
i do think @fbhutto is totally \\
& & wannabeish... :) \\ \hline
@Cluster-2-user-1 & @Cluster-2-user-2 & Yeah, I don't like her either. I like
Imran Khan better. \\ \hline
@Cluster-2-user-2 & @Cluster-2-user-1 & i liked him till i saw a report ver his
old 92 world cup winning team \\
& & members said they never liked him.. \\ \hline
@Cluster-2-user-2 & @Cluster-2-user-1 & 								
if he culdnt get his own team members behind him then.....
\\ \hline 	
@Cluster-2-user-1 & @Cluster-2-user-2 & Well, can't say anything about that. But
he's a good person, a \\
& & better one in politics at least. \\ \hline
@Cluster-2-user-2 & @Cluster-2-user-1 & well..sub continental politics na...everyone
has a murky past or \\
& & present...just depends on how deep one digs..no offence intended
\\ \hline 						
@Cluster-2-user-1 & @Cluster-2-user-2 &
None taken. But look at his hospital. Despite a few controversies, I \\
& & can tell you firsthand that it's amazing. Good man.	 \\ \hline
@Cluster-2-user-1 & &
\#becauseoftwitter I've become more vocal in terms of venting. \\ \hline 		
@Cluster-2-user-1 & & \#becauseoftwitter I've become more sensitive. I lose a
follower and \\ & & I go berserk.	\\ \hline \hline
\end{tabular}
\end{center}
\end{table}

\newpage
\subsection*{Table 13 - Sample Activity Listing}
Sample activity of a typical user in Cluster 3.

\begin{table}[htb!]
\label{typical_act_cluster3}
\begin{center}
\begin{tabular}{|l|l|l|}
\hline
Posted by & Intended for & Text content \\ \hline \hline
@Cluster-3-user & & My maid told today; somebody threw one day old daughter
wrapped \\
& & in a polythene bag in Filth Drum outside a
house. \#humanitarian \\ \hline
@Cluster-3-user & & I'll write on this very issue; daughter, poverty or an
illegitimate child \\
& & was she? why our social system allows us to take these plunges?
\\ \hline 							
@Friend-1 & @Cluster-3-user & did the baby survived? what has went wrong
with people, they are \\
& & becoming so heartless. :( \\ \hline
@Cluster-3-user & @Friend-1 & haan she survived and somebody has adopted
her.. only there.. but it \\
& & is heart breaking..	
\\ \hline
@Friend-1 & @Cluster-3-user & oh Thanks God. Yes it is really heart
breaking. \\ \hline
@Friend-2 & @Cluster-3-user & alive ???? =O \\ \hline
@Cluster-3-user & @Friend-2 & yeah alive.. an issueless couple has
adopted her \\ \hline
@Friend-2 & @Cluster-3-user & Thank God! .... thts literally inhuman
act! I mean the adopting parents \\
& & should atleast get back to her parents and hang them!\\ \hline 				
@Cluster-3-user & @Friend-2 & where would they find them? to hang \\ \hline
@Cluster-3-user & & Half of Pakistanis say match fixing allegations against
cricketers untrue \\
& & [URL] \#Cricket \#Pakistan \\ \hline
@Friend-3 & @Cluster-3-user & which means half of Pakistan believes the
allegations of \#Cricket \\
& & \#MatchFixing against the \#Pakistan team to be true \#Honesty \#ICC \\ \hline 				
@Cluster-3-user & @Friend-3 & ha ha ha yeah.. agreed \\ \hline
@Friend-4 & @Cluster-3-user & The proper way of match fixing is thru umpires,
didnt give collingwood \\
& & out he made a big partnership \& gave well set akmal out \\ \hline
@Cluster-3-user & @Friend-4 & true.. collingwood has fixed match.. LOL \\ \hline
\hline
\end{tabular}
\end{center}
\end{table}


\ignore{
\section*{Additional Files}
  \subsection*{Additional file 1 --- Sample additional file title}
    Additional file descriptions text (including details of how to
    view the file, if it is in a non-standard format or the file extension).  This might
    refer to a multi-page table or a figure.

  \subsection*{Additional file 2 --- Sample additional file title}
    Additional file descriptions text.
}

\end{bmcformat}
\end{document}